\documentclass[preprint,12pt]{elsarticle}
\usepackage[utf8]{inputenc}
\usepackage{tikz}
\usepackage{booktabs}
\usepackage{siunitx}
\usepackage{makecell}
\usetikzlibrary{arrows.meta, positioning, calc, shapes.geometric}
\usepackage{subcaption}  % add to preamble
\usepackage{psfrag} % for \psfrag command
\usepackage{algpseudocode}
\usepackage{algorithmicx}
\usepackage{algcompatible}
\usepackage{algorithm}
\usepackage{amssymb}
\usepackage{graphicx}
\usepackage{bbm}
\usepackage{amsmath}
\usepackage{xcolor}
\usepackage{caption}
\usepackage{url} 
\usepackage{ifthen}

\journal{XXXXXXX}

\begin{document}

\begin{frontmatter}

%% Title, authors and addresses

%% use the tnoteref command within \title for footnotes;
%% use the tnotetext command for theassociated footnote;
%% use the fnref command within \author or \affiliation for footnotes;
%% use the fntext command for theassociated footnote;
%% use the corref command within \author for corresponding author footnotes;
%% use the cortext command for theassociated footnote;
%% use the ead command for the email address,
%% and the form \ead[url] for the home page:
%% \title{Title\tnoteref{label1}}
%% \tnotetext[label1]{}
%% \author{Name\corref{cor1}\fnref{label2}}
%% \ead{email address}
%% \ead[url]{home page}
%% \fntext[label2]{}
%% \cortext[cor1]{}
%% \affiliation{organization={},
%%             addressline={},
%%             city={},
%%             postcode={},
%%             state={},
%%             country={}}
%% \fntext[label3]{}

\title{Unlocking feedforward capabilities in Model Predictive Control algorithms to deal with measurable disturbances}

%% use optional labels to link authors explicitly to addresses:
 \author[1]{José Luis Guzmán*}
 \author[1]{Igor Pataro}
 \author[1]{Juan D. Gil}
 \author[1]{Manuel Berenguel}

%% Author affiliation
\address[1]{University of Almería, Department of Informatics, CIESOL, ceiA3, La Cañada de San Urbano s/n, Almería, Spain. \{joseluis.guzman@ual.es\}}

%% Abstract
\begin{abstract}
%% Text of abstract
Disturbance rejection is a central objective in process control, particularly when measurable disturbances can be exploited through feedforward action. Although Model Predictive Control (MPC) naturally incorporates disturbance models and prediction capabilities, standard formulations cannot achieve complete disturbance rejection since the cost function penalises control effort. This limitation prevents MPC from reproducing the behaviour of classical feedforward compensators. This work proposes a novel framework to embed true feedforward capabilities within MPC without removing the control effort penalty. The approach introduces a dual-control structure in which two control actions are computed simultaneously: a tracking-oriented action addressing set-point tracking and robustness, and a feedforward-oriented action dedicated to disturbance rejection. Both contributions are combined into a single control signal on which the process constraints are explicitly enforced. The feedforward-oriented action is formulated without penalising control effort, enabling full compensation of measurable disturbances. The methodology is developed for Dynamic Matrix Control (DMC), Generalised Predictive Control (GPC), and state-space MPC. Its effectiveness is demonstrated through simulation studies, including comparisons with standard MPC and classical feedforward schemes. A case study based on a reverse osmosis process shows that the proposed approach improves disturbance rejection while preserving constraint handling and overall control performance.
\end{abstract}

%%Graphical abstract
%\begin{graphicalabstract}
%\includegraphics{grabs}
%\end{graphicalabstract}

%%Research highlights
% \begin{highlights}
% \item Research highlight 1
% \item Research highlight 2
% \end{highlights}

%% Keywords
\begin{keyword}
%% keywords here, in the form: keyword \sep keyword
Feedforward control \sep MPC \sep Disturbances \sep GPC \sep DMC
%% PACS codes here, in the form: \PACS code \sep code

%% MSC codes here, in the form: \MSC code \sep code
%% or \MSC[2008] code \sep code (2000 is the default)

\end{keyword}

\end{frontmatter}

%% Add \usepackage{lineno} before \begin{document} and uncomment 
%% following line to enable line numbers
%% \linenumbers

%% main text
%%

%% Use \section commands to start a section

\section{Introduction}

Disturbance rejection, or the regulation problem, is one of the most fundamental challenges in control engineering, particularly in the process industry. In a regulation control problem, the objective is to maintain a process variable close to a desired setpoint despite disturbances that tend to drive the system away from this operating condition \citep{hagglund2023process}.

When disturbance information is not available, that is, when disturbances are non-measurable, the rejection problem must be addressed through feedback control strategies. In such cases, the controller responds to deviations in the process variable after the disturbance has already affected the system, thereby attenuating its impact. In contrast, when disturbances are measurable, their information can be exploited to anticipate their effect on the process. This enables compensation before the disturbance affects the controlled variable, potentially eliminating or significantly reducing its impact. This is the fundamental principle of feedforward control \citep{guzman2024feedforward}. 

Feedforward is a proactive control strategy that compensates for measurable disturbances before any response is observed in the process output. It is not a replacement for feedback control, but rather complements it. In the ideal unconstrained case, when the disturbance model is perfectly known, and no model-inversion issues arise, feedforward control can completely cancel the disturbance's effect. However, in practical scenarios where inversion is not feasible, or the model is imperfect, tuning rules for the feedforward compensator must be employed \citep{guzman2024feedforward}. The combination of feedforward and feedback control allows measurable disturbances to be counteracted proactively, while feedback mechanisms handle unmeasurable disturbances, setpoint tracking, and model uncertainties \citep{guzman2024feedforward}. This combined feedforward-feedback structure is widely used in industrial practice and effectively decouples the control objectives of disturbance rejection and set-point tracking.

%In the ideal case, when the disturbance model is perfectly known and no model-inversion issues arise, feedforward control can completely cancel the disturbance's effect. However, in practical scenarios where inversion is not feasible, or the model is imperfect, feedforward compensator tuning rules from the literature must be employed \citep{guzman2024feedforward}. This combined feedforward-feedback structure is widely used in industrial practice, where feedforward control is typically integrated with PID control loops. In such configurations, feedforward compensates for measurable disturbances, while feedback ensures robustness against uncertainties and handles residual errors. This effectively decouples the different control objectives for disturbance rejection and set-point tracking.

Feedforward capabilities have traditionally been incorporated into Model Predictive Control (MPC) algorithms, as disturbance models can be explicitly included in the output predictions. By accounting for these models, MPC naturally extends its predictive and constraint-handling capabilities to disturbance compensation problems. In particular, when future disturbance values are available or can be reliably estimated, MPC can anticipate their effects in advance, which is especially valuable in process control applications~\citep{camacho2026mpc,li2026model}.

Nevertheless, despite these advantages, standard MPC formulations are generally unable to completely reject measurable disturbances. Even under ideal conditions, where the disturbance model is perfectly known, the disturbance signal is fully available, no model inversion issues arise, and the input is not saturated, the resulting control action typically only attenuates the disturbance effect. This limitation stems from the structure of the MPC objective function, which penalises control effort, thereby reducing the aggressive input actions required for complete rejection and inherently enforcing a trade-off between performance and actuation effort. Moreover, MPC computes control trajectories that simultaneously address multiple objectives, including setpoint tracking, disturbance rejection, and control effort minimisation. Consequently, the conflicting nature of these objectives limits the ability of standard MPC formulations to replicate the behaviour of a pure feedforward compensator.

The disturbance rejection properties of MPC have been studied, for instance, in the context of Generalised Predictive Control (GPC), where it has been shown that achieving pure feedforward behaviour requires the control effort weighting term in the cost function to be set to zero \citep{pawlowski2012improving,PAWLOWSKI20171239}. Similar results have been reported for the Infinite Horizon MPC (IHMPC) \citep{PATARO2022b} and for the Linear Quadratic optimal controller with FF (LQ-FF) \citep{Pataro2023i}. As seen, removing the control effort penalty can recover feedforward-like behaviour, but at the expense of excessively high bandwidth, increased sensitivity to model uncertainty, and aggressive setpoint-tracking responses \cite{PAWLOWSKI20171239}. To mitigate these issues, in \cite{PAWLOWSKI20171239}, a reference filter was added to the original GPC algorithm and a modified GPC structure based on the Filtered Smith Predictor was proposed to improve robustness.

%The presented results in the literature are consistent with classical feedforward control principles. To fully cancel the effect of a disturbance, the controller often must apply aggressive control actions to anticipate its impact. In MPC, however, such aggressive actions are penalized through the control effort term in the cost function, which inherently enforces a trade-off between performance and actuation effort. As a result, the control actions required for perfect disturbance rejection are suppressed, preventing complete compensation.

%Moreover, MPC computes control trajectories that simultaneously address multiple objectives, including setpoint tracking, disturbance rejection, and control effort minimization. Consequently, the conflicting nature of these objectives limits the ability of standard MPC formulations to replicate the behavior of a pure feedforward compensator. As commented above, in \citep{PAWLOWSKI20171239} was shown that removing the control effort penalty can recover feedforward-like behavior, but at the expense of excessively high bandwidth, increased sensitivity to model uncertainty, and aggressive responses in setpoint tracking. To mitigate these issues, a reference filter was added to the original GPC algorithm and a modified GPC structure based on the Filtered Smith Predictor was proposed to improve robustness.

The aforementioned limitations, together with the need for additional control structures within the MPC framework, motivate the development of predictive control formulations that explicitly reproduce feedforward behaviour while preserving robustness and reasonable control effort within a unified framework. To this end, this work proposes a novel approach to systematically incorporate true feedforward capabilities into MPC without altering the nominal feedback design. 

The proposed approach is inspired by classical feedforward-feedback control structures, in which feedforward and feedback actions are computed separately and then combined into the process control action, with the feedforward controller acting in open loop. Accordingly, the MPC algorithm is extended to handle two different control components: one associated with the standard MPC objectives (e.g., tracking and robustness), and a second open-loop component dedicated exclusively to disturbance rejection. The latter is designed without penalising control effort, thereby enabling complete disturbance rejection. Both control signals are integrated into a unified optimisation framework that preserves MPC's constraint-handling capabilities while embedding pure feedforward behaviour. In this work, the proposed formulation is developed and analysed under nominal conditions to clearly establish and demonstrate the methodology. Therefore, the focus is on recovering feedforward disturbance-rejection capabilities within MPC rather than on providing a formal analysis of closed-loop stability or robustness properties, which remain important topics for future research.

Related ideas have also been explored in the literature to enhance the set-point-tracking capabilities of predictive controllers via feedforward-like mechanisms. In \citep{carrasco_feedforward_2011,goodwin_preview_2011}, additional feedforward structures were incorporated into MPC formulations to improve set-point tracking performance by anticipating reference changes and reducing transient errors. These approaches share the common philosophy of decomposing the control action into complementary components with different objectives. However, their primary focus is on enhancing reference tracking rather than on measurable disturbance rejection. In contrast, the present work addresses the fundamentally different problem of measurable disturbance compensation through feedforward action. As in classical control theory, feedforward compensation for set-point tracking and measurable disturbance rejection constitute distinct control problems and require different formulations and design considerations. While previous predictive feedforward approaches focused on improving servo performance, the proposed methodology is designed to recover the pure disturbance-rejection capabilities of classical feedforward compensators within the MPC framework, while preserving constraint handling and robustness. To the best of the authors’ knowledge, this constitutes a novel formulation for systematically embedding true feedforward disturbance compensation into MPC without modifying the nominal feedback controller design.

The proposed methodology is derived for three widely used MPC formulations: Dynamic Matrix Control (DMC), Generalised Predictive Control (GPC), and state-space MPC, covering algorithms based on convolution, transfer function, and state-space models. The formulation is intentionally developed for the nominal case to highlight the underlying principles of the proposed feedforward embedding methodology. Moreover, the algorithm is derived for linear systems which allows to apply the superposition principle. Its effectiveness is demonstrated through simulation studies comparing the proposed approach with conventional MPC algorithms and classical feedforward schemes. Additionally, a comprehensive case study involving a reverse osmosis process is presented.

The remainder of this paper is organised as follows. Section~\ref{sec:preliminaries} summarises the main concepts of classical feedforward control and the basic formulation of MPC algorithms, including the treatment of measurable disturbances in the prediction model. Section~\ref{sec:extended_mpc} presents the proposed extended MPC formulation with embedded feedforward capabilities and discusses its relationship with classical feedforward compensation. Section~\ref{sec:simulation} provides the simulation studies, including a comparative analysis and the reverse osmosis case study. Finally, Section~\ref{sec:conclusions} draws the main conclusions and outlines future research directions.

\section{Preliminaries}
\label{sec:preliminaries}

This section briefly describes the basic information on a classical feedforward control scheme, the common features of MPC algorithms, such as the cost function and the calculation of the control law considering the disturbance model, and the derivation of the specific prediction equations for the DMC, GPC, and state-space MPC controllers.

\subsection{Feedforward control}

\begin{figure}[h]
    \centering
    \includegraphics[width=0.9\linewidth]{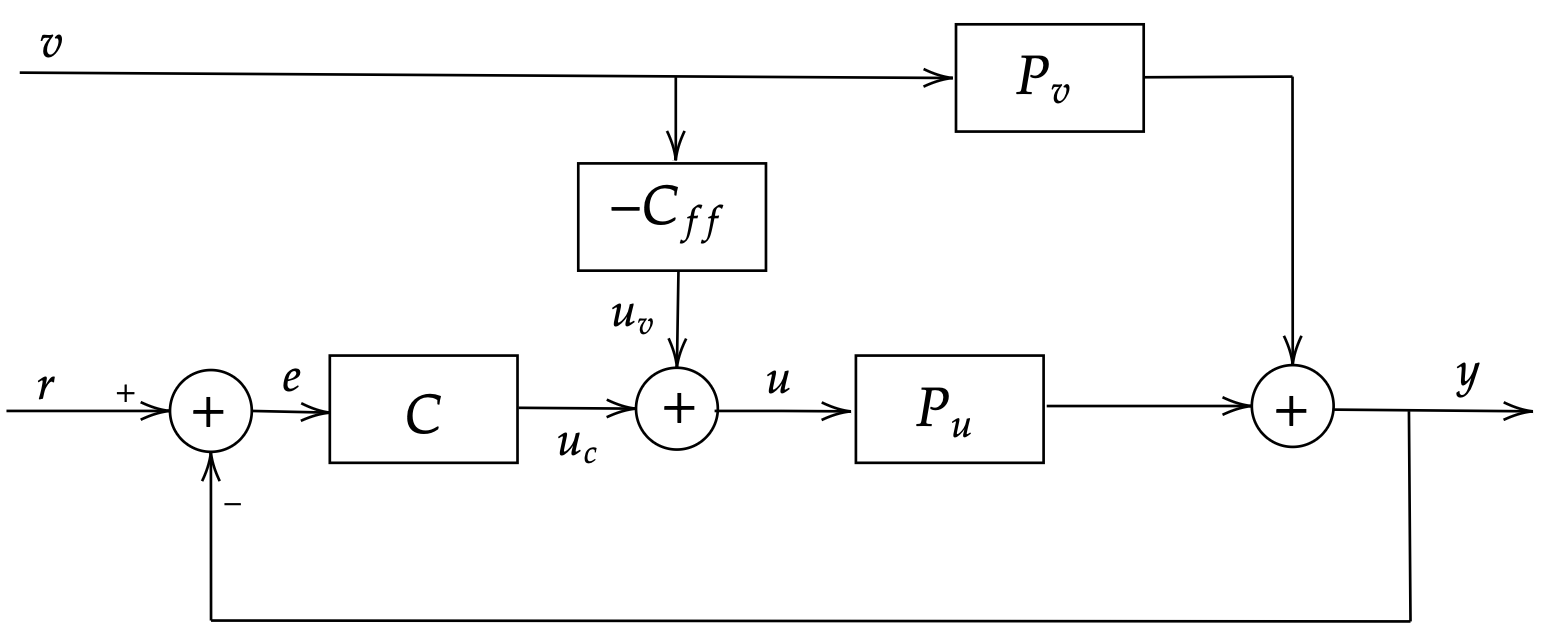}
    \caption{Classical feedforward control scheme to deal with measurable disturbances.}
    \label{fig:classical_ff}
\end{figure}

Feedforward control can be considered the classical solution for addressing the measurable disturbance-rejection problem and mitigating the limitations of feedback control described earlier. The classical feedforward control scheme is presented in Figure~\ref{fig:classical_ff}, where $C$ is the feedback controller, $P_u$ represents the process dynamics from the manipulated variable to the process output, and $r$, $u$, and $y$ denote the set-point, the total control signal, and the process output, respectively. Moreover, a measurable load disturbance $v$ affects the process output through the disturbance dynamics $P_v$. The feedforward compensator is denoted by $C_{ff}$ and acts in open loop, using only measurable disturbance information to anticipate and counteract the effect of $v$ before it propagates to the process output.

In this structure, the total control signal applied to the process is obtained as the combination of two contributions:
\begin{equation}
u(t)=u_c(t)+u_v(t),
\label{eq:u_feedback_feedforward}
\end{equation}
where $u_c(t)$ is the feedback control action generated by controller $C$, and $u_v(t)$ is the feedforward control action generated from the measurable disturbance through $C_{ff}$. Thus, feedback and feedforward actions are designed with different objectives and subsequently combined at the process input.

%Feedforward control can be considered the classical solution for addressing the measurable disturbance-rejection problem and mitigating the limitations of feedback control described earlier. The classical feedforward control scheme is presented in Figure \ref{fig:classical_ff}, where $C$ is the feedback controller, signals $r$, $u$, and $y$ are the set-point, the control signal, and the process output, respectively, and $P_u$ represents the process dynamics relating the control signal with the process output. Moreover, a load disturbance $v$ influences the feedback loop according to the figure, with transfer function $P_v$ between load $v$ and process output $y$. The feedforward compensator is represented by $C_{ff}$ and is connected in open-loop to counteract the effect of the measurable disturbance.

According to this scheme, the transfer function between disturbance $v$ and process output $y$  becomes
\begin{equation}
G_{y/v}=\frac{P_v-P_uC_{ff}}{1+CP_u},
\label{eq:cl_y_d_ff}
\end{equation}

\noindent where the feedforward compensator should be calculated in the following way in order to remove the disturbance effect:
\begin{equation}
C_{ff}=\frac{P_v}{P_u}.
\label{eq:ff_design}
\end{equation}

When \eqref{eq:ff_design} is realisable and is substituted in \eqref{eq:cl_y_d_ff}, the disturbance signal is completely removed before affecting the process output. Otherwise, alternative control schemes or tuning rules for the feedforward compensator should be used to account for the non-realisable problems \citep{guzman2024feedforward}.

%Notice that the motivation of this work relies on modifying the MPC control algorithms to include the capabilities of the control scheme of Figure \ref{fig:classical_ff} to perfectly reject disturbance signals in the perfect cancellation scenario. 

This classical structure provides the conceptual basis for the methodology proposed in this work. The main objective is to extend standard MPC algorithms so that they reproduce the decoupled feedback-feedforward architecture in Figure~\ref{fig:classical_ff}: a feedback-oriented predictive action devoted to set-point tracking, robustness, and constraint handling, and an open-loop feedforward-oriented predictive action devoted to measurable disturbance rejection.

\subsection{Model predictive control}

MPC algorithms compute the control action by solving, at each sampling instant, a finite-horizon optimisation problem that explicitly accounts for system dynamics and measurable disturbances \citep{camacho2026mpc}. The common objective in MPC is to determine a sequence of future control increments, $\Delta u(t+j-1)$, that minimises a multistage cost function of a SISO (Single Input Single Output) system in the form
\begin{equation}
J=\sum_{j=N_{1}}^{N}\delta(j)[\hat{y}(t+j|t)-w(t+j)]^2+\sum_{j=1}^{N_{u}}\lambda(j)[\Delta
u(t+j-1)]^2, \label{eq_1}
\end{equation}

\noindent where $\hat{y}(t+j|t)$ denotes the predicted system output at time $t+j$ based on information available at time $t$ (here denoted in discrete-time domain), and $\Delta u(t+j-1)$ represents the future control increments, with $\Delta = (1 - z^{-1})$. The parameters $N_1$ and $N$ define the minimum and maximum prediction horizons, respectively, while $N_u$ is the control horizon. The weighting sequences $\delta(j)$ and $\lambda(j)$ penalise tracking errors and control effort over the prediction horizon and constitute key tuning parameters of the controller. The reference trajectory $w(t+j)$ may correspond to the setpoint or to a filtered version of it, providing a smooth transition from the current output $y(t)$ to the desired reference \citep{camacho2026mpc}. Note that a SISO model is used for simplifying the algorithm presentation; for MIMO (Multi-Input Multi-Output) systems, the formulation can be straightforwardly extended \citep{camacho2026mpc}.

%\noindent where $\hat{y}(k+j|t)$ is an optimal system output prediction sequence performed with data known up to discrete time $t$, $\Delta u(t+j-1)$ is a future control increment sequence obtained from cost function minimization with $\Delta=(1-z^{-1})$, $N_1$ and $N$ are the minimum and maximum prediction horizons respectively, $N_{u}$ is the control horizon and $\delta(j)$ and $\lambda(j)$ are weighting sequences that penalize the future tracking errors and control efforts, respectively, along the horizons. The horizons and weighting sequences are design parameters used as tuning knobs. The reference trajectory $w(k+j)$ can be the set-point or a smooth approximation from the current value of the system output $y(t)$ towards the known reference by means of a determined filter \citep{camacho2026mpc}.

In \eqref{eq_1}, the $j$-step-ahead prediction of the system output, $\hat{y}(t+j|t)$, is generated using a process model that depends on the specific MPC formulation (e.g., transfer function, state-space, finite impulse response, etc.), as described in the following subsections. For a prediction horizon of length $N$ and a control horizon $N_u$, the sequence of predicted outputs can be expressed in a compact form as a function of past inputs and outputs, future control increments, and measurable disturbances. This leads to the following general prediction equation:

\begin{equation}
\begin{aligned}
\underbrace{
\begin{bmatrix}
\hat{y}(t+d+1|t) \\
\hat{y}(t+d+2|t) \\
\vdots \\
\hat{y}(t+d+N|t)
\end{bmatrix}}_{\mathbf{\widehat{y}}}
= &
\underbrace{
\begin{bmatrix}
g_1 & 0 & \cdots & 0 \\
g_2 & g_1 & \cdots & 0 \\
\vdots & \vdots & \ddots & \vdots \\
g_{N} & g_{N-1} & \cdots & g_{N-N_u}
\end{bmatrix}}_{\mathbf{G}}
\underbrace{
\begin{bmatrix}
\Delta u(t) \\
\Delta u(t+1) \\
\vdots \\
\Delta u(t+N_u-1)
\end{bmatrix}}_{\mathbf{u}}
\\ 
& + \underbrace{
\begin{bmatrix}
h_1 & 0 & \cdots & 0 \\
h_2 & h_1 & \cdots & 0 \\
\vdots & \vdots & \ddots & \vdots \\
h_{N} & h_{N-1} & \cdots & h_1
\end{bmatrix}}_{\mathbf{H}}
\underbrace{
\begin{bmatrix}
\Delta v(t+1) \\
\Delta v(t+2) \\
\vdots \\
\Delta v(t+N)
\end{bmatrix}}_{\mathbf{v}}
 + \underbrace{
\begin{bmatrix}
f_1 \\
f_2 \\
\vdots \\
f_N
\end{bmatrix}}_{\mathbf{f}},
\end{aligned}
\label{eq:prediction_general}
\end{equation}

\noindent where $g_1,\ldots,g_{N-N_u}$ in $\mathbf{G}$ denote the step response coefficients of the process related to the input, $\mathbf{H}$ is the matrix containing the step response coefficients associated with the disturbance, and $f_1,\ldots,f_N$ represent the free response of the system, which depends on past inputs, outputs, and disturbances. 

On the other hand, $\mathbf{u}$ includes the future control signals (decision variables), and the vector $\mathbf{v}$ gathers the future disturbance increments. In some applications, particularly when disturbances are related to process loads, future disturbance values may be known a priori or estimated using trends or forecasting techniques. In such cases, the contribution of the disturbance to the predicted output can be explicitly computed through the second term in \eqref{eq:prediction_general}. 

Conversely, if the disturbance is assumed to remain constant over the prediction horizon, i.e., $v(t+j)=v(t)$ $\forall j =[1, N] $, then $\Delta v(t+j)=0$, and the future disturbance contribution vanishes.

The prediction model in \eqref{eq:prediction_general} can be written in the compact form
\begin{equation}
\mathbf{\widehat{y}}= \mathbf{G}\mathbf{u}+\mathbf{H}\mathbf{v}+\mathbf{f},
\label{eq:compact_prediction}
\end{equation}

\noindent where $\mathbf{f}$ represents the free response of the system, which depends on past inputs, outputs, and disturbances, and current measurements of outputs and disturbances. $\mathbf{G}\mathbf{u}$ is the forced response depending of the future control actions $\mathbf{u}$, and the term $\mathbf{H}\mathbf{v}$ accounts for the contribution of future measurable disturbances to the predicted output. 

The specific structure of $\mathbf{f}$, as well as the matrices involved in \eqref{eq:compact_prediction}, depends on the adopted prediction model and will be detailed for each MPC formulation in the following subsections.

The optimal control sequence, $\mathbf{u}$, is obtained by substituting the prediction model \eqref{eq:compact_prediction} into the cost function \eqref{eq_1} and expressing it in a compact quadratic form. After straightforward algebraic manipulation, the cost function can be written as
\begin{equation}
\begin{aligned}
J = & \mathbf{u}^\top (\mathbf{Q}_{\lambda} + \mathbf{G}^\top \mathbf{G}) \mathbf{u} 
+ 2(\mathbf{H}\mathbf{v} + \mathbf{f} - \mathbf{w})^\top \mathbf{G}\mathbf{u}, \\ 
& + (\mathbf{H}\mathbf{v} + \mathbf{f} - \mathbf{w})^\top (\mathbf{H}\mathbf{v} + \mathbf{f} - \mathbf{w}),
\end{aligned}
\label{eq:J_2}
\end{equation}

\noindent where $\mathbf{w} = [w(t+d+1),\, w(t+d+2),\, \hdots,\, w(t+d+N)]^\top$ denotes the reference trajectory over the prediction horizon, and $\mathbf{Q}_{\lambda}$ is a diagonal weighting matrix that penalizes the control increments, with elements $\lambda(j)$. The tracking weights $\delta(j)$ do not explicitly appear in \eqref{eq:J_2} because they are typically chosen as constant along the prediction horizon, i.e., $\delta(j)=1$. Under this common assumption, the control weighting sequence is also taken to be constant over the horizon N, $\lambda(j)=\lambda$, yielding $\mathbf{Q}_{\lambda} = \lambda \mathbf{I}$. In this case, the controller tuning is effectively reduced to selecting the scalar parameter $\lambda$, which balances tracking performance and control effort, and is often normalised by the process gain.

In an unconstrained problem, the optimal control sequence is obtained by minimising \eqref{eq:J_2} with respect to $\mathbf{u}$. Since the cost function is quadratic, the solution can be derived analytically by setting the gradient with respect to $\mathbf{u}$ equal to zero, leading to the well-known least-squares solution:
\begin{equation}
\mathbf{u}=\mathbf{K}^{-1}\mathbf{G}^\top(\mathbf{w}-\mathbf{H}\mathbf{v}-\mathbf{f}),
\label{c_l_1}
\end{equation}

\noindent where $\mathbf{K}=\mathbf{Q}_{\lambda}+\mathbf{G}^\top\mathbf{G}$.

According to the receding horizon strategy, only the first element of the optimal control sequence is applied to the process, i.e., $\Delta u(t)$. Denoting by $\mathbf{k}$ the first row of $\mathbf{K}^{-1}$, the implemented control action is given by
\begin{equation}
\Delta u(t)=\mathbf{k}\mathbf{G}^\top(\mathbf{w}-\mathbf{H}\mathbf{v}-\mathbf{f}).
\label{c_l}
\end{equation}

Equation~\eqref{c_l} corresponds to the well-known unconstrained MPC control law, explicitly accounting for future reference trajectories and measurable disturbances. It is worth noting that the contribution of the disturbance term is filtered through the matrix $\mathbf{K}^{-1}\mathbf{G}^\top$, which depends on the control weighting. As a result, the disturbance compensation is inherently coupled with the control effort penalisation, limiting the ability of standard MPC formulations to achieve complete disturbance rejection \citep{pawlowski2012improving}.

The previous optimal control action is common for most linear MPC frameworks. The main difference among MPC algorithms lies in how the prediction equation \eqref{eq:compact_prediction} is formulated and the free response term $\mathbf{f}$ is obtained. This will be briefly summarised in the following subsections for GPC, DMC and state space MPC algorithms. More details can be found in \citep{camacho2026mpc}.

\subsubsection{Prediction equation for DMC}

In DMC, the prediction model is constructed directly from the process step responses and the measurable disturbance. Let $\{g_i\}_{i=1}^{\infty}$ and $\{h_i\}_{i=1}^{\infty}$ denote the step response coefficients relating the manipulated input increments $\Delta u$ and the measurable disturbance increments $\Delta v$ to the output, respectively. Then, the predicted output $j$ steps ahead can be written as
\begin{equation}
\hat{y}(t+j|t)=\sum_{i=1}^{\infty} g_{i} \Delta u(t+j-i)+\sum_{i=1}^{\infty} h_{i} \Delta v(t+j-i)+\hat{n}(t+j|t),
\label{eq:dmc_pred0}
\end{equation}
where $\hat{n}(t+j|t)$ accounts for unmeasured disturbances and modelling errors.

Assuming, as in standard DMC, that $\hat{n}(t+j|t)=\hat{n}(t|t)=y(t)-\hat{y}(t|t)$ remains constant over the prediction horizon (with $y(t)$ the actual measured process output), and separating future and past increments, the prediction can be rearranged as
\begin{equation}
\hat{y}(t+j|t)=\sum_{i=1}^{j} g_{i} \Delta u(t+j-i)+\sum_{i=1}^{j} h_{i} \Delta v(t+j-i+1)+f(t+j),
\label{eq:dmc_pred1}
\end{equation}
where $f(t+j)$ is the free response, i.e., the part of the prediction that does not depend on future control or disturbance increments. So, the free response is given by
\begin{equation}
f_{DMC}(t+j)=y(t)+\sum_{i=1}^{\infty}(g_{j+i}-g_i)\Delta u(t-i)+\sum_{i=1}^{\infty}(h_{j+i}-h_i)\Delta v(t-i+1).
\label{eq:dmc_free_scalar}
\end{equation}
%\noindent in which $y(t)$ is the output measured value.

For asymptotically stable systems, -the step response coefficients converge to constant values, so the previous expression can be truncated to the finite step response length. Stacking the predictions over the horizon yields
\begin{equation}
\mathbf{\widehat{y}_{DMC}}=\mathbf{G}\mathbf{u}+\mathbf{H}\mathbf{v}+\mathbf{f_{DMC}}.
\end{equation}
 
The free response vector is then written as

\begin{equation}
\mathbf{f_{DMC}}=\mathbf{G}_p\mathbf{u}_p +\mathbf{H}_p \mathbf{v}_p + \mathbf{L}y(t),
\label{eq:f_dmc}
\end{equation}
where $\mathbf{u}_p$ and $\mathbf{v}_p$ collect the past control and disturbance increments, respectively, while $\mathbf{G}_p$, $\mathbf{H}_p$, and $\mathbf{L}$ are matrices built from the process and disturbance step response coefficients.

Therefore, in DMC, the free response is explicitly expressed in terms of the measured current output and the past increments in control and disturbance. The detailed formulation can be followed in \citep{camacho2026mpc}.

\subsubsection{Prediction equation for the GPC}

In GPC, the prediction equation is derived from the CARIMA model
\begin{equation}
A(z^{-1})y(t)=z^{-d}B(z^{-1})\Delta u(t-1)+D(z^{-1})\Delta v(t)+\frac{\xi(t)}{\Delta},
\label{eq:carima_gpc}
\end{equation}
where $A(z^{-1})=1+a_1 z^{-1}+\cdots+a_{n_a}z^{-n_a}$ is the autoregressive polynomial, $B(z^{-1})=b_0+b_1 z^{-1}+\cdots+b_{n_b}z^{-n_b}$ is the input polynomial, and $D(z^{-1})=d_0+d_1 z^{-1}+\cdots+d_{n_d}z^{-n_d}$ represents the disturbance dynamics. The operator $z^{-1}$ denotes the backward shift operator (unit delay), and $\xi(t)$ is a zero-mean disturbance term.

By solving the corresponding Diophantine equations, the predicted output is decomposed into a forced response, driven by future control actions, and a free response, which depends only on information available up to time $t$. Accordingly, the stacked prediction over the horizon can be written as
\begin{equation}
\mathbf{\widehat{y}_{GPC}}=\mathbf{G}\mathbf{u}+\mathbf{H}\mathbf{v}+\mathbf{f_{GPC}},
\end{equation}
where the free response vector is given by
\begin{equation}
\mathbf{f_{GPC}}=\mathbf{G}_{p}\mathbf{u}_p+\mathbf{H}_{p}\mathbf{v}_p+\mathbf{S}\mathbf{y}_p,
\label{eq:f_gpc}
\end{equation}

\noindent where $\mathbf{G}_{p}$, $\mathbf{H}_{p}$, and $\mathbf{S}$ are matrices constructed from the model polynomials and the Diophantine solutions, accounting for the effects of past control increments, past disturbance increments, and past outputs, respectively. The vectors $\mathbf{u}_p$, $\mathbf{v}_p$, and $\mathbf{y}_p$ collect the relevant past values of the control increments, disturbance increments, and system output, typically organised as
\begin{equation}
\begin{aligned}
    \mathbf{u}_p = & [\Delta u(t-1),\, \Delta u(t-2),\, \ldots]^\top,\\
    \mathbf{v}_p = & [\Delta v(t),\, \Delta v(t-1),\, \ldots]^\top,\\
    \mathbf{y}_p = & [y(t),\, y(t-1),\, \ldots]^\top.
\end{aligned}
\end{equation}

Hence, in the GPC formulation, the free response corresponds to the component of the predicted output fully determined by past data. Further details of the obtained matrices can be found in \citep{camacho2026mpc}.

\subsubsection{Prediction equation for state-space MPC}

In this work, the state-space MPC is derived from an augmented model formulated in incremental form in order to be equivalent to the external representation forms of the DMC and GPC. Starting from the original discrete-time system, the state vector is extended to implicitly account for input increments and transport delays associated with both manipulated inputs and measurable disturbances.

The augmented state vector is defined as
\begin{equation}
\mathbf{x_a}(t)=\begin{bmatrix} \mathbf{x}(t) \\ \mathbf{z}_u(t) \\ \mathbf{z}_d(t) \\ u_{\text{prev}}(t) \\ d_{\text{prev}}(t) \end{bmatrix},
\label{eq:xa_ss}
\end{equation}
where $\mathbf{x}(t)$ are the original process states, $\mathbf{z}_u(t)$ and $\mathbf{z}_d(t)$ are the delay-chain states associated with manipulated inputs and disturbances, respectively, and $u_{\text{prev}}(t)$ and $d_{\text{prev}}(t)$ store the previous values of the inputs and disturbances required to recover the incremental signals, that is $u_{\text{prev}}(t) = u(t-1)$, and $d_{\text{prev}}(t) = d(t-1)$.

The augmented system can then be written as
\begin{equation}
\mathbf{x_a}(t+1)= \mathbf{A_a} \mathbf{x_a}(t)+\mathbf{B_a} \Delta u(t)+\mathbf{B_{d,a}}\Delta v(t),
\end{equation}
\begin{equation}
y(t)=\mathbf{C_a} \mathbf{x_a}(t),
\end{equation}
where the matrices $\mathbf{A_a}$, $\mathbf{B_a}$, $\mathbf{B_{d,a}}$, and $\mathbf{C_a}$ are constructed from the original system matrices and the delay-chain dynamics. In particular, $\mathbf{A_a}$ contains the original state dynamics, the shift structure of the delay chains, and the update equations for the stored input and disturbance values. The matrices $\mathbf{B_a}$ and $\mathbf{B_{d,a}}$ map the control and disturbance increments into both the original states and the delay-chain states, while $\mathbf{C_a}$ selects the output from the augmented state only related to the system dynamics, not to the auxiliary states.

Stacking the predictions over the horizon leads to
\begin{equation}
\mathbf{\widehat{y}_{SS}}=\mathbf{G}\mathbf{u}+\mathbf{H}\mathbf{v} + \mathbf{f_{SS}},
\end{equation}
where 
\begin{equation}
\mathbf{f_{SS}}=\mathbf{\Phi}\mathbf{x_a}(t),
\label{eq:f_ss}
\end{equation}
being $\mathbf{\Phi}$ the free-response matrix, obtained from the augmented matrices as

\begin{equation}
\mathbf{\Phi}=
\begin{bmatrix}
\mathbf{C_a} \mathbf{A_a}\\
\mathbf{C_a} \mathbf{A_a}^2 \\
\vdots \\
\mathbf{C_a} \mathbf{A_a}^N
\end{bmatrix}.
\end{equation}

Hence, in the state-space formulation, the free response is compactly expressed as a function of the current augmented state. Importantly, although written in terms of $\mathbf{x_a}(t)$, this term implicitly contains the effects of past inputs, past disturbances, and their associated delays, since all these quantities are embedded in the augmented state vector. A detailed description of the matrices and augmented states can be found in \cite{PATARO2022b}.

\subsubsection{Unified prediction model summary}

As shown in the previous subsections, although DMC, GPC, and state-space MPC formulations are derived from different modelling paradigms, namely step response, transfer function, and state-space representations, respectively, all prediction models can ultimately be expressed using the same compact structure introduced in \eqref{eq:compact_prediction}.

The term $\mathbf{G}\mathbf{u}$ represents the forced response associated with future control actions and is common to all MPC formulations. Similarly, the term $\mathbf{H}\mathbf{v}$ accounts for the contribution of future measurable disturbances whenever future disturbance information is available. The main difference among MPC algorithms lies in the formulation of the free response term $\mathbf{f}$, which encapsulates the effect of past information and current system measurements according to the adopted prediction model.

Table~\ref{tab:prediction_summary} summarises the prediction equations derived for the considered MPC formulations and highlights the specific structure of the free response component in each case.

\begin{table}[h]
\centering
\caption{Summary of prediction equations for different MPC formulations.}
\label{tab:prediction_summary}
\renewcommand{\arraystretch}{1.4}
\begin{tabular}{p{2.2cm} p{4cm} p{6.5cm}}
\hline
\textbf{MPC} & \textbf{Prediction model} & \textbf{Free response term} \\
\hline

DMC &
$\mathbf{\widehat{y}}=
\mathbf{G}\mathbf{u}
+
\mathbf{H}\mathbf{v}
+
\mathbf{f_{DMC}}$ &
$\mathbf{f_{DMC}} = \mathbf{G}_{p}\mathbf{u}_{p} +\mathbf{H}_{p}\mathbf{v}_{p} +\mathbf{L}y(t)$ \\

GPC &
$\mathbf{\widehat{y}}=
\mathbf{G}\mathbf{u}
+
\mathbf{H}\mathbf{v}
+
\mathbf{f_{GPC}}$ &
$\mathbf{f_{GPC}}
=
\mathbf{G}_{p}\mathbf{u}_{p}
+
\mathbf{H}_{p}\mathbf{v}_{p}
+
\mathbf{S}\mathbf{y}_{p}$ \\

SS MPC &
$\mathbf{\widehat{y}}=
\mathbf{G}\mathbf{u}
+
\mathbf{H}\mathbf{v}
+
\mathbf{f_{SS}}$ &
$\mathbf{f_{SS}}
=
\mathbf{\Phi\mathbf{}x_a}(t)$ \\

\hline
\end{tabular}
\end{table}

It is important to note that despite apparent differences in the free-response formulation, all approaches implicitly account for the effects of past control actions, disturbances, delays, and current process information. Therefore, the methodology proposed in the next section can be formulated in a unified manner, independent of the adopted MPC framework, by exploiting the common compact prediction structure.

\section{Extended MPC formulation with feedforward capabilities}
\label{sec:extended_mpc}

As highlighted in the previous section, despite the different modelling paradigms used in DMC, GPC, and state-space MPC formulations, all prediction models can ultimately be expressed through the same compact structure. This unified prediction structure enables a general methodology for extending MPC capabilities, independent of the specific predictive model employed. Since measurable disturbances are already explicitly incorporated into the prediction equation, future disturbance information can naturally be accounted for within the optimisation problem. However, despite these predictive capabilities, complete rejection of measurable disturbances cannot generally be achieved. This limitation does not arise from the prediction model itself, but rather from the inherent structure of the MPC objective function, where control effort penalisation introduces a trade-off between control performance and actuator activity (see Equation~\eqref{eq:J_2}). Consequently, the aggressive control actions typically required to fully compensate for measurable disturbances are attenuated by the optimisation process, preventing standard MPC formulations from reproducing the behaviour of classical feedforward compensators.

Motivated by these observations, this section presents a unified extension of MPC algorithms aimed at systematically embedding feedforward capabilities while preserving the nominal feedback-oriented MPC design. The proposed methodology exploits the common prediction structure previously identified and can therefore be formulated independently of the adopted modelling framework. First, a general formulation is introduced based on the common compact prediction model. Subsequently, the methodology is particularised for the different MPC algorithms presented in the previous section, namely DMC, GPC, and state-space MPC, highlighting the specific modifications required in each case.

\subsection{Proposed methodology}

Inspired by the classical feedforward control principles presented in Section~\ref{sec:preliminaries}, the proposed framework systematically embeds pure feedforward capabilities into MPC while preserving the standard control effort penalisation associated with feedback-oriented objectives. The methodology reproduces the structure of a classical feedforward-feedback controller, where set-point tracking and measurable disturbance rejection are treated as decoupled control problems and addressed through separate optimal control actions.

More specifically, the control problem is decomposed into two complementary components. A first control action is dedicated to conventional MPC objectives, including setpoint tracking and robustness. A second control action is explicitly devoted to disturbance rejection and is formulated without penalising control effort, thereby recovering the anticipative behaviour required for complete feedforward compensation. Both control actions are computed simultaneously and subsequently combined to generate the final manipulated variable applied to the process. Unlike classical feedforward schemes, however, the predictive nature of MPC enables both control components to be integrated into a unified optimisation framework with intrinsic constraint handling capabilities. This feature enables explicit accounting for actuator limitations and interactions between feedback and feedforward actions, particularly in situations where feedforward inversion issues arise, as previously demonstrated in classical feedforward compensation schemes \citep{guzman2024feedforward}.

%In order to incorporate feedforward capabilities within the MPC framework, the future system response is decomposed into two components. The first component represents the nominal system evolution without the effect of measurable disturbances, while the second component captures the contribution of measurable disturbances and enables feedforward compensation.

Let the predicted output sequence be expressed as two independent contributions:

\begin{equation}
\mathbf{\widehat{y}}_c = \mathbf{G}\mathbf{u}_c + \mathbf{f}_c,
\label{eq:control}
\end{equation}
\begin{equation}
\mathbf{\widehat{y}}_v = \mathbf{G}\mathbf{u}_v + \mathbf{f}_v.
\label{eq:ff}
\end{equation}

Here, $\mathbf{\widehat{y}}_c \in \mathbb{R}^{N}$ represents the nominal predicted response without disturbance effects, whereas $\mathbf{\widehat{y}}_v \in \mathbb{R}^{N}$ accounts for the predicted effect of measurable disturbances. The control sequences $\mathbf{u}_c, \mathbf{u}_v \in \mathbb{R}^{N_u}$ correspond to the nominal MPC action for set-point tracking and feedback control purposes, and the feedforward-oriented MPC action for disturbance rejection, respectively. The vectors $\mathbf{f}_c$ and $\mathbf{f}_v \in \mathbb{R}^{N}$ represent the corresponding free responses, without and with measurable disturbance information, respectively. Notice that $\mathbf{f}_c$ incorporates feedback information about the current state of the plant through measured outputs and past control actions, thereby supporting closed-loop predictive control objectives. In contrast, $\mathbf{f}_v$ only includes the contribution of the process model together with past and future measurable disturbance information, without incorporating feedback information from the plant. Consequently, $\mathbf{u}_v$ is computed as an open-loop feedforward-oriented control action exclusively devoted to disturbance compensation.

To obtain a compact formulation, the control variables are grouped into a single augmented vector:
\begin{equation}
\mathbf{U} =
\begin{bmatrix}
\mathbf{u}_c \\
\mathbf{u}_v
\end{bmatrix},
\label{eq:uext}
\end{equation}

\noindent and the block-diagonal dynamic matrix is defined as:
\begin{equation}
\mathbf{G}_u =
\begin{bmatrix}
\mathbf{G} & \mathbf{0} \\
\mathbf{0} & \mathbf{G}
\end{bmatrix}.
\label{eq:Gext}
\end{equation}

Similarly, define the reference and free response mismatch vector:
\begin{equation}
\mathbf{F} =
\begin{bmatrix}
\mathbf{f}_c - \mathbf{w}_c \\
\mathbf{f}_v - \mathbf{w}_v
\end{bmatrix},
\label{eq:fext}
\end{equation}

\noindent where $\mathbf{w}_c$ and $\mathbf{w}_v$ are the reference trajectories associated with each component.

The proposed cost function is defined as:
\begin{equation}
J =
(\mathbf{\widehat{y}}_c - \mathbf{w}_c)^{T}(\mathbf{\widehat{y}}_c - \mathbf{w}_c)
+
(\mathbf{\widehat{y}}_v - \mathbf{w}_v)^{T}(\mathbf{\widehat{y}}_v - \mathbf{w}_v)
+
\mathbf{u}_c^{T}\mathbf{Q}_{\lambda_c}\mathbf{u}_c
+
\mathbf{u}_v^{T}\mathbf{Q}_{\lambda_v}\mathbf{u}_v,
\label{eq:newJ}
\end{equation}

\noindent where $\mathbf{Q}_{\lambda_c}$ and $\mathbf{Q}_{\lambda_v}$ are weighting matrices associated with the control effort of each component. In particular, $\mathbf{Q}_{\lambda_v}$ can be set to zero in order to recover pure feedforward behavior, as will be described later.

Substituting the prediction models (\ref{eq:control}) and (\ref{eq:ff}) in (\ref{eq:newJ}), the cost function becomes:
\begin{align}
J = &
(\mathbf{G}\mathbf{u}_c + \mathbf{f}_c - \mathbf{w}_c)^T
(\mathbf{G}\mathbf{u}_c + \mathbf{f}_c - \mathbf{w}_c) \nonumber \\
& +
(\mathbf{G}\mathbf{u}_v + \mathbf{f}_v - \mathbf{w}_v)^T
(\mathbf{G}\mathbf{u}_v + \mathbf{f}_v - \mathbf{w}_v) \nonumber \\
& +
\mathbf{u}_c^T \mathbf{Q}_{\lambda_c} \mathbf{u}_c
+
\mathbf{u}_v^T \mathbf{Q}_{\lambda_v} \mathbf{u}_v,
\end{align}

This expression can be rewritten in compact quadratic form using \eqref{eq:uext}, \eqref{eq:Gext}, and \eqref{eq:fext} as:
\begin{equation}
J(\mathbf{U}) =
\mathbf{U}^T (\mathbf{G}_u^T \mathbf{G}_u + \mathbf{\Lambda}) \mathbf{U}
+
2 \mathbf{U}^T \mathbf{G}_u^T \mathbf{F}
+
\mathbf{F}^T \mathbf{F},
\end{equation}

\noindent where
\begin{equation}
\mathbf{\Lambda} =
\begin{bmatrix}
\mathbf{Q}_{\lambda_c} & \mathbf{0} \\
\mathbf{0} & \mathbf{Q}_{\lambda_v}
\end{bmatrix}.
\end{equation}

The optimal solution when no constraints are considered is obtained by minimizing $J(\mathbf{U})$ with respect to $\mathbf{U}$:
\begin{equation}
\frac{\partial J}{\partial \mathbf{U}} =
2(\mathbf{G}_u^T \mathbf{G}_u + \mathbf{\Lambda})\mathbf{U}
+
2\mathbf{G}_u^T \mathbf{F} = 0,
\end{equation}

which leads to the analytical solution:
\begin{equation}
\mathbf{U}^{\star} =
-(\mathbf{G}_u^T \mathbf{G}_u + \mathbf{\Lambda})^{-1}
\mathbf{G}_u^T \mathbf{F}.
\end{equation}

Then, using again the expressions \eqref{eq:uext}, \eqref{eq:Gext}, and \eqref{eq:fext}, we can get 
\begin{equation}
\mathbf{U}^{\star}
=
\begin{bmatrix}
\mathbf{u}_c^{\star} \\
\mathbf{u}_v^{\star}
\end{bmatrix}
=
\begin{bmatrix}
(\mathbf{G}^T\mathbf{G}+\mathbf{Q}_{\lambda_c})^{-1} & \mathbf{0} \\
\mathbf{0} & (\mathbf{G}^T\mathbf{G}+\mathbf{Q}_{\lambda_v})^{-1},
\end{bmatrix}
\begin{bmatrix}
\mathbf{G}^T(\mathbf{w}_c-\mathbf{f}_c) \\
\mathbf{G}^T(\mathbf{w}_v-\mathbf{f}_v)
\end{bmatrix}.
\end{equation}

Thus,
\begin{equation}
\mathbf{u}_c^{\star} = \mathbf{K}_c(\mathbf{w}_c-\mathbf{f}_c),
\end{equation}
\begin{equation}
\mathbf{u}_v^{\star} = \mathbf{K}_v(\mathbf{w}_v-\mathbf{f}_v),
\label{eq:uv_calc}
\end{equation}

with 

\begin{equation}
\mathbf{K}_c = (\mathbf{G}^T\mathbf{G}+\mathbf{Q}_{\lambda_c})^{-1}\mathbf{G}^T,
\end{equation}
\begin{equation}
\mathbf{K}_v = (\mathbf{G}^T\mathbf{G}+\mathbf{Q}_{\lambda_v})^{-1}\mathbf{G}^T.
\label{eq:Kv_calc}
\end{equation}

Considering the receding horizon strategy, only the first elements of the optimal control sequences are applied to the process, namely $\Delta u_c(t)$ and $\Delta u_v(t)$. Following the classical feedforward-feedback architecture introduced in Section~\ref{sec:preliminaries}, and particularly the control structure described by \eqref{eq:u_feedback_feedforward}, the final manipulated variable is obtained as the combination of two independently computed control contributions: a feedback-oriented predictive action and a feedforward-oriented predictive action. Therefore, the total control increment is given by
\begin{equation}
\Delta u(t)
=
\Delta u_c(t)
+
\Delta u_v(t).
\label{eq:totalu}
\end{equation}

In this way, the proposed formulation explicitly reproduces the decoupled feedback-feedforward structure of classical feedforward compensation (see Equation~(\ref{eq:u_feedback_feedforward})), while embedding both actions into a unified MPC optimisation framework. This formulation preserves the standard MPC capabilities for set-point tracking through $\mathbf{u}_c$, while enabling explicit feedforward compensation through $\mathbf{u}_v$ without penalising control effort. This result is particularly relevant, as it enables the set-point tracking and measurable disturbance rejection problems to be explicitly decoupled and independently designed within the same MPC framework, which is not possible using conventional MPC formulations. Consequently, the proposed methodology overcomes one of the main structural limitations of standard MPC algorithms, where both objectives are intrinsically coupled through a single cost function and a common control action.

Although the proposed formulation enables an explicit decoupling between the set-point tracking and disturbance rejection problems, complete disturbance compensation is not automatically guaranteed yet. As previously discussed, pure feedforward behaviour requires aggressive control actions that are typically attenuated by the control effort penalisation embedded in the MPC objective function. Therefore, the feedforward-oriented component must be specifically designed to recover the behaviour of a classical feedforward compensator.

Following the ideas presented in \citep{pawlowski2012improving,PAWLOWSKI20171239}, pure feedforward disturbance rejection can be recovered by formulating the feedforward-oriented control problem as a regulation-to-zero problem and eliminating the control effort penalisation associated with this component. To this end, the reference trajectory associated with the disturbance compensation problem is set to zero, that is, $\mathbf{w}_v = \mathbf{0}$, while the corresponding control effort weighting matrix is removed, $\mathbf{Q}_{\lambda_v} = \mathbf{0}$.

Under these assumptions, the feedforward-oriented control action given by (\ref{eq:uv_calc}) and (\ref{eq:Kv_calc}) becomes
\begin{equation}
\mathbf{u}_v^{\star}
=
-\mathbf{K}_v \mathbf{f}_v
=
-(\mathbf{G}^T\mathbf{G})^{-1}\mathbf{G}^T \mathbf{f}_v,
\label{eq:uv_lambda0}
\end{equation}

\noindent where
\begin{equation}
\mathbf{K}_v=
(\mathbf{G}^T\mathbf{G})^{-1}\mathbf{G}^T
\end{equation}

\noindent corresponds to the least-squares inverse of the dynamic matrix. In the particular case where $\mathbf{G}$ is square and nonsingular, the previous expression reduces to
\begin{equation}
\mathbf{u}_v^{\star}
=
-\mathbf{G}^{-1}\mathbf{f}_v,
\label{eq:uv_final}
\end{equation}
which explicitly reveals the classical feedforward compensation structure based on process inversion, as will be presented in more detail in the next subsections. This ideal inversion case is assumed throughout the present work for further analysis and to better show the proposed methodology.

As can be observed, the feedforward-oriented component is exclusively driven by the process model together with past and predicted measurable disturbance information embedded in $\mathbf{f}_v$, while no feedback information from the plant is incorporated into its formulation. Consequently, the resulting control action is computed in open loop and no longer depends on tracking objectives or control effort minimisation. This behaviour directly reproduces the anticipative nature of a classical feedforward compensator, while remaining systematically integrated within the MPC framework.

Considering the formulation of the free response terms for the different MPC algorithms described in the previous section in (\ref{eq:uv_final}), it can be verified that the proposed feedforward-oriented component naturally recovers a structure analogous to that of a classical feedforward compensator. In particular, substituting the corresponding free-response expression into the feedforward control law yields a formulation that explicitly separates the process dynamics from the disturbance dynamics.

In contrast, the feedback MPC component preserves the conventional formulation:
\begin{equation}
\mathbf{u}_c^{\star}
=
\mathbf{K}_c(\mathbf{w}_c-\mathbf{f}_c),
\end{equation}
thereby retaining the closed-loop properties associated with set-point tracking, robustness, and constraint handling. The final manipulated variable is therefore obtained by combining a feedback-oriented MPC action and a pure feedforward compensation term.

Notice that another important advantage of the proposed methodology is that, unlike classical feedforward compensation schemes, the intrinsic constraint handling capabilities of MPC can be naturally exploited to manage the interaction between the feedback and feedforward control actions in a systematic and coupled manner. As presented previously in \eqref{eq:totalu}, both control components are combined to generate the final manipulated variable. Consequently, actuator constraints can be directly imposed on the resulting control action, allowing both control components to cooperate while respecting process limitations. For instance, input saturation constraints can be formulated as
\begin{equation}
u_{\min}
\leq
u(t-1)+\Delta u_c(t)+\Delta u_v(t)
\leq
u_{\max}.
\end{equation}

Similarly, slew rate constraints can be incorporated as
\begin{equation}
\Delta u_{\min}
\leq
\Delta u_c(t)+\Delta u_v(t)
\leq
\Delta u_{\max}.
\end{equation}

These constraints are imposed on the combined control action rather than on each individual component, allowing the optimisation process to systematically coordinate the contribution of the feedback and feedforward terms along the prediction horizon. This feature is particularly relevant in situations where feedforward inversion problems arise or aggressive feedforward actions may drive the manipulated variable towards saturation. In such cases, the proposed MPC formulation naturally resolves the interaction between both control objectives while preserving feasibility and respecting actuator limitations.

Moreover, beyond input magnitude and slew rate limitations, any classical MPC constraint can be naturally incorporated into the proposed methodology by formulating it over the combined feedback-feedforward control action. This includes, for instance, output constraints, soft constraints, terminal constraints, state constraints in state-space formulations, and operating region limitations commonly encountered in process control applications \citep{camacho2026mpc}.

\subsection{Application to DMC}

For the DMC formulation, the proposed methodology presented in the previous section is obtained by decomposing the free response into a feedback-oriented component and a feedforward-oriented component according to
\begin{eqnarray}
\mathbf{f}_c &=& \mathbf{G}_{p}\mathbf{u}_{p,c}
+\mathbf{L}y(t), \\
\mathbf{f}_v &=&
\mathbf{G}_{p}\mathbf{u}_{p,v}
+\mathbf{L}\hat{y}(t)
+\mathbf{H}\mathbf{v}
+\mathbf{H}_{p}\mathbf{v}_{p},
\label{eq:fv_dmc}
\end{eqnarray}

\noindent where the feedforward free response, $\mathbf{f}_v$, is constructed exclusively from the process model and measurable disturbance information, without incorporating plant feedback measurements. Hence, $\hat{y}(t)$ denotes the current model-based output prediction replacing the measured process output in the feedforward path.

Substituting \eqref{eq:fv_dmc} into the feedforward control law \eqref{eq:uv_final} leads to

\begin{equation}
\mathbf{u}_v^{\star}
=
-\mathbf{G}^{-1}
\mathbf{G}_{p}\mathbf{u}_{p,v}
-\mathbf{G}^{-1}
\mathbf{L}\hat{y}(t)
-
\underbrace{
\mathbf{G}^{-1}\mathbf{H}\mathbf{v}
-
\mathbf{G}^{-1}\mathbf{H}_{p}\mathbf{v}_{p}
}_{\text{feedforward}},
\label{eq:dmc_ff}
\end{equation}

\noindent where the last two terms explicitly recover the structure of a classical feedforward compensator. The disturbance-to-output dynamics, represented by $\mathbf{H}$ and $\mathbf{H}_p$, are effectively compensated through inversion of the process dynamics represented by $\mathbf{G}$. Meanwhile, the first two terms account for the internal evolution of the feedforward action, including the contribution of past feedforward control moves and the model-based prediction.

When future disturbance information is unavailable, the preview term vanishes, i.e., $\mathbf{v}=\mathbf{0}$, and only the contribution associated with $\mathbf{v}_p$ remains. In this situation, the proposed feedforward-oriented action becomes equivalent to a classical feedforward compensator relying solely on the currently available measurable disturbance signal.

\subsection{Application to GPC}

The extension to the GPC algorithm follows the same rationale. In this case, the feedback-oriented and feedforward-oriented free responses become
\begin{eqnarray}
\mathbf{f}_c &=&
\mathbf{G}_{p}\mathbf{u}_{p,c}
+\mathbf{S}\mathbf{y}_{p}, \\
\mathbf{f}_v &=&
\mathbf{G}_{p}\mathbf{u}_{p,v}
+\mathbf{S}\mathbf{\hat{y}}_{p}
+\mathbf{H}\mathbf{v}
+\mathbf{H}_{p}\mathbf{v}_{p},
\label{eq:fv_gpc}
\end{eqnarray}

\noindent where the measured output sequence $\mathbf{y}_p$ is replaced in the feedforward path by the model-based prediction sequence $\mathbf{\hat{y}}_p$, thereby preserving the open-loop nature of the feedforward component.

Substituting \eqref{eq:fv_gpc} into \eqref{eq:uv_lambda0} yields

\begin{equation}
\mathbf{u}_v^{\star}
=
-\mathbf{G}^{-1}
\mathbf{G}_{p}\mathbf{u}_{p,v}
-\mathbf{G}^{-1}
\mathbf{S}\mathbf{\hat{y}}_{p}
-
\underbrace{
\mathbf{G}^{-1}\mathbf{H}\mathbf{v}
+
\mathbf{G}^{-1}\mathbf{H}_{p}\mathbf{v}_{p}
}_{\text{feedforward}},
\label{eq:gpc_ff}
\end{equation}

\noindent where the same feedforward compensation structure identified for DMC is recovered. The last two terms represent the inversion of the disturbance-to-output dynamics through the process dynamics, whereas the first two terms account for the internal dynamic contribution of the feedforward-oriented controller based on the past information and model-based prediction.

Likewise, when future disturbance information is unavailable, the preview contribution vanishes, and the resulting formulation reduces to a feedforward compensator driven only by the currently measurable disturbance signal.

\subsection{Application to the state-space MPC}

Assuming that the system states are completely observable, when the proposed methodology is applied to the state-space MPC formulation, the free responses $\mathbf{f}_c$ and $\mathbf{f}_v$ are obtained from the augmented state model as
\begin{eqnarray}
\mathbf{f}_c &=& \mathbf{\Phi}\mathbf{x}_{a,c}(t), \\
\mathbf{f}_v &=& \mathbf{\Phi}\mathbf{\hat{x}}_{a,v}(t)
+
\mathbf{H}\mathbf{v},
\label{eq:fv_ss}
\end{eqnarray}

\noindent where $\mathbf{x}_{a,c}(t)$ corresponds to the augmented state vector employed by the feedback-oriented MPC component, constructed from the current plant measurements, while $\mathbf{\hat{x}}_{a,v}(t)$ represents the augmented state used by the feedforward component. The latter is propagated using the process model and measurable disturbance information only, without incorporating feedback information from the plant output. In this way, the feedforward component preserves its open-loop nature while still accounting for the dynamic evolution of the process.

Substituting (\ref{eq:fv_ss}) into the feedforward control law (\ref{eq:uv_lambda0}) yields

\begin{equation}
\mathbf{u}_v^{\star}
=
-\mathbf{G}^{-1}
\mathbf{\Phi}\mathbf{\hat{x}}_{a,v}(t)
-
\underbrace{
\mathbf{G}^{-1}
\mathbf{H}\mathbf{v}
}_{\text{feedforward}},
\label{eq:ss_ff}
\end{equation}

\noindent where a classical feedforward compensation structure can again be identified. The future contribution of the measurable disturbance, represented by $\mathbf{H}\mathbf{v}$, is effectively compensated through inversion of the control-to-output dynamics represented by $\mathbf{G}$, with opposite sign. This reproduces the same fundamental structure observed in conventional feedforward compensators based on process inversion.

In the state-space formulation, the effect of past feedforward control actions, measurable disturbances, and the internal process model estimation is implicitly embedded in the augmented feedforward state vector, $\mathbf{\hat{x}}_{a,v}(t)$, according to its definition in \eqref{eq:xa_ss}. Consequently, the information required to sustain the feedforward action over time is internally accessible through the state representation, without requiring explicit terms associated with past control increments or disturbance values, as occurs in DMC and GPC formulations. Therefore, the feedforward control component remains entirely determined by the process model and measurable disturbance information, while preserving its open-loop nature and avoiding the incorporation of plant feedback measurements.

Moreover, when future disturbance information is unavailable, the disturbance preview vector becomes $\mathbf{v}=\mathbf{0}$. In this situation, the feedforward-oriented MPC action is determined by the internal augmented feedforward state, $\mathbf{\hat{x}}_{a,v}(t)$, which already contains the effects of past disturbance values and feedforward control actions. Hence, the resulting behaviour becomes equivalent to that of a classical feedforward compensator operating without disturbance preview, where only the currently available measurable disturbance information is exploited.

\section{Simulation study}
\label{sec:simulation}

This section presents two simulation studies to show the capabilities of the proposed methodology and to compare it with conventional MPC algorithms and classical feedforward schemes. First, a linear example is selected to illustrate the main advantages in perfect cancellation, constraint handling, and inversion problems. After that, a nonlinear process using a reverse osmosis unit is selected to evaluate the algorithms beyond nominal linear conditions. Since all MPC formulations presented in this paper yield equivalent nominal responses under the same prediction model and tuning conditions, the simulation results in this section are based on the GPC algorithm as a representative benchmark. The corresponding implementations for the three MPC formulations are publicly available on GitHub \citep{guzman2026efmpc}, enabling full reproducibility of the results and facilitating further comparative studies.

The performance of the different control strategies was evaluated using the Integral Absolute Error (IAE) according to:

\begin{equation}
IAE = \sum_{k=1}^{N} |r(k)-y(k)|.
\end{equation}

In order to provide a more detailed analysis of the different operating conditions considered in the dynamic simulations, four different IAE indices were computed. Specifically, IAE$_{\mathrm{Ref}}$ evaluates the response associated with the set-point tracking task, IAE$_{\mathrm{Dis}}$ quantifies the measurable disturbance rejection performance, IAE$_{\mathrm{Mix}}$ evaluates the simultaneous disturbance rejection and set-point tracking scenario, respectively, and IAE$_{\mathrm{Tot}}$ represents the overall accumulated error over the complete simulation horizon.

\subsection{Linear example}

In order to evaluate the proposed methodology under different feedforward compensation conditions, a comparative simulation study for a linear process is used considering three representative scenarios: (i) an ideal measurable disturbance rejection scenario, (ii) an input-constrained feedforward compensation scenario, and (iii) a non-invertible feedforward compensation scenario. The objective of these case studies is to analyse the ability of the proposed formulation to recover classical feedforward behaviour under ideal conditions, as well as to assess its performance when practical limitations such as actuator constraints and feedforward inversion problems are present. For all scenarios, the same dynamic simulation sequence was considered in order to facilitate a consistent comparison among the different control strategies. The simulations were performed over 180 sampling instants. A set-point change was introduced at time instant 30, followed by a measurable disturbance at instant 60. Subsequently, a second measurable disturbance was applied at instant 130 together with an additional set-point change, thereby creating a more demanding operating condition where disturbance rejection and set-point tracking objectives must be simultaneously addressed.

Four different control strategies were evaluated and compared: (i) a conventional GPC controller combined with an external classical feedforward compensator (denoted as External FF), using a non-aggressive controller tuning with control weighting factor $\lambda = 1$; (ii) a classical intrinsic feedforward GPC formulation with the same controller tuning, i.e., $\lambda = 1$,  (denoted as Internal FF); (iii) the same intrinsic feedforward GPC formulation using an aggressive controller tuning with $\lambda=0$ (referred to as Internal FF ($\lambda=0$)); and (iv) the proposed extended feedforward GPC formulation (denoted as EF-GPC), where the feedback-oriented component employed the same non-aggressive tuning configuration with $\lambda_c = 1$, while the feedforward-oriented component was configured with $\lambda_v = 0$ in order to recover pure feedforward compensation behaviour. For all predictive controllers, a sampling time of $T_s=1~\mathrm{s}$ was adopted. The prediction and control horizons were selected as $N=N_u=60$.
%It should be noted that the EF-GPC algorithm has been used as the reference formulation throughout the simulation studies. Nevertheless, the obtained results and conclusions are directly extensible to DMC and state-space MPC formulations, since the proposed methodology is derived from the common compact prediction structure shared by all these predictive control approaches, as presented above.

\subsubsection{Ideal measurable disturbance rejection scenario}

The first considered case corresponds to the ideal measurable disturbance rejection scenario, where perfect feedforward compensation is theoretically achievable. To this end, the process and disturbance dynamics were selected such that no model inversion problems, non-minimum phase dynamics, or infeasible delay relationships are present. The process and disturbance transfer functions are given by:
\begin{align}
    P_u(s) &= \frac{1}{10s + 1} e^{-10s}, \\
    P_v(s) &= \frac{0.8}{5s + 1} e^{-15s},
\end{align}

\noindent where the disturbance transfer function exhibits a larger delay than the manipulated-input transfer function, thereby allowing the feedforward compensator to be implemented in a realizable manner according to classical feedforward control theory. 

\begin{figure}[htbp]
    \centering
\includegraphics[width=\linewidth]{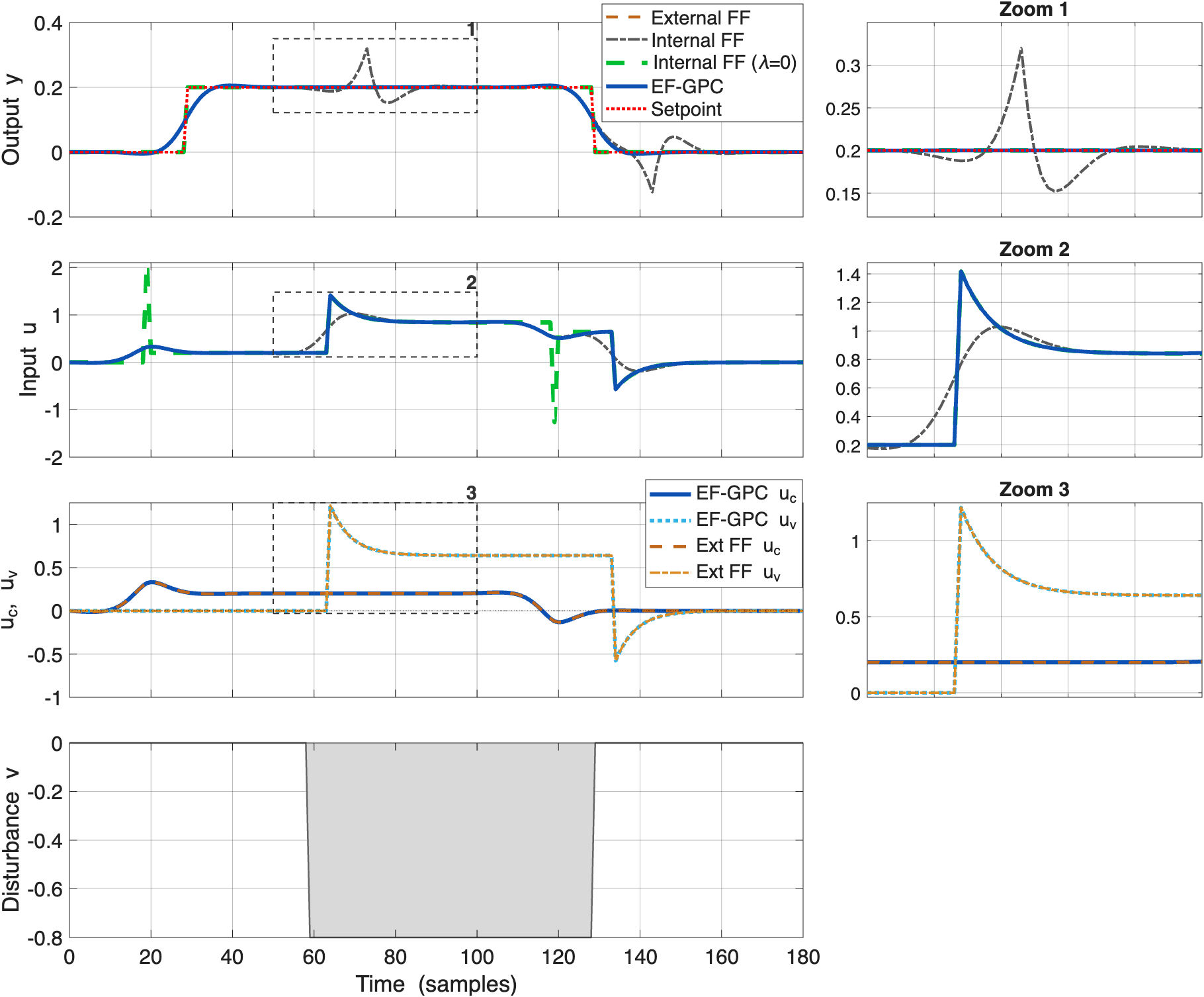}
    \caption{Comparative simulation results for the ideal measurable disturbance rejection scenario}
    \label{fig:ideal_scenario}
\end{figure}

The results obtained from the corresponding simulation study are presented in Figure~\ref{fig:ideal_scenario}. As observed, the External FF, Internal FF, and EF-GPC strategies exhibit identical responses during the first set-point change. This behaviour is expected, as all these controllers use the same non-aggressive tuning configuration. In contrast, the Internal FF ($\lambda=0$) strategy achieves perfect set-point tracking due to the absence of control-effort penalisation, although at the expense of a significantly more aggressive control action, making this configuration unsuitable for most practical implementations given actuator and operational constraints.

Regarding disturbance rejection, around simulation instant 60, it can be observed that all strategies achieve perfect disturbance compensation in this ideal scenario except the conventional Internal FF formulation (see Zoom 1, Zoom 2, and Zoom 3). In this case, the use of a common tuning parameter for both the set-point tracking and disturbance rejection objectives introduces a compromise between both control tasks, thereby deteriorating the disturbance rejection performance.

Finally, when a measurable disturbance and a set-point change are simultaneously introduced at instant 130, the proposed EF-GPC formulation preserves the same perfect disturbance rejection and set-point tracking performance achieved by the External FF strategy. This result highlights the ability of the proposed methodology to explicitly decouple both control objectives while maintaining the nominal MPC tuning for the feedback-oriented component.

The quantitative performance indices associated with the previous simulation study are summarized in Table~\ref{tab:ff_comparison}. As expected from the responses shown in Figure~\ref{fig:ideal_scenario}, the proposed EF-GPC formulation achieves exactly the same performance indices as the classical External FF strategy in all evaluated operating conditions.  In contrast, the conventional Internal FF strategy presents a significant degradation in the disturbance rejection performance, leading to considerably larger values of IAE$_{\mathrm{Dis}}$, IAE$_{\mathrm{Mix}}$, and consequently IAE$_{\mathrm{Tot}}$. This behaviour is directly associated with the intrinsic compromise introduced by the use of a single tuning parameter to simultaneously address tracking and disturbance rejection objectives. Finally, the Internal FF ($\lambda=0$) configuration  achieves zero error in all scenarios due to the absence of control penalisation. However, this behaviour is obtained at the expense of highly aggressive control actions and reduced robustness margins, making such tuning impractical for real applications.

\begin{table}[ht]
\centering
\caption{Comparison of performance indices for the ideal measurable disturbance rejection scenario.}
\label{tab:ff_comparison}

\renewcommand{\arraystretch}{1.2}

\begin{tabular}{lcccc}
\toprule
\textbf{Strategy} &
\textbf{IAE$_{\mathrm{Ref}}$} &
\textbf{IAE$_{\mathrm{Dis}}$} &
\textbf{IAE$_{\mathrm{Mix}}$} &
\textbf{IAE$_{\mathrm{Total}}$} \\
\midrule
EF-GPC                    & 0.230 & 0.000 & 0.230 & 0.460 \\
External FF               & 0.230 & 0.000 & 0.230 & 0.460  \\
Internal FF               & 0.246 & 0.635 & 0.980 & 1.861  \\
Internal FF ($\lambda=0$) & 0.000 & 0.000 & 0.000 & 0.000 \\
\bottomrule
\end{tabular}

\end{table}

Overall, these results demonstrate that the proposed EF-GPC formulation is capable of reproducing the same ideal disturbance rejection performance achieved by a classical external feedforward compensator used in industry, while simultaneously preserving a smooth set-point tracking behaviour. This result is particularly significant since the proposed methodology is directly inspired by the classical feedforward-feedback control structure, but explicitly decouples the tracking and disturbance rejection objectives within the predictive control framework.

\subsubsection{Input-constrained feedforward compensation scenario}

The second considered case analyses the behaviour of the different control strategies when actuator limitations are explicitly taken into account. To this end, the same process and disturbance dynamics considered in the previous subsection were retained, while an input saturation constraint was incorporated into the control problem according to:

\begin{equation}
-1.3 \leq u(t) \leq 1.3.
\end{equation}

\begin{figure}[h]
    \centering
\includegraphics[width=\linewidth]{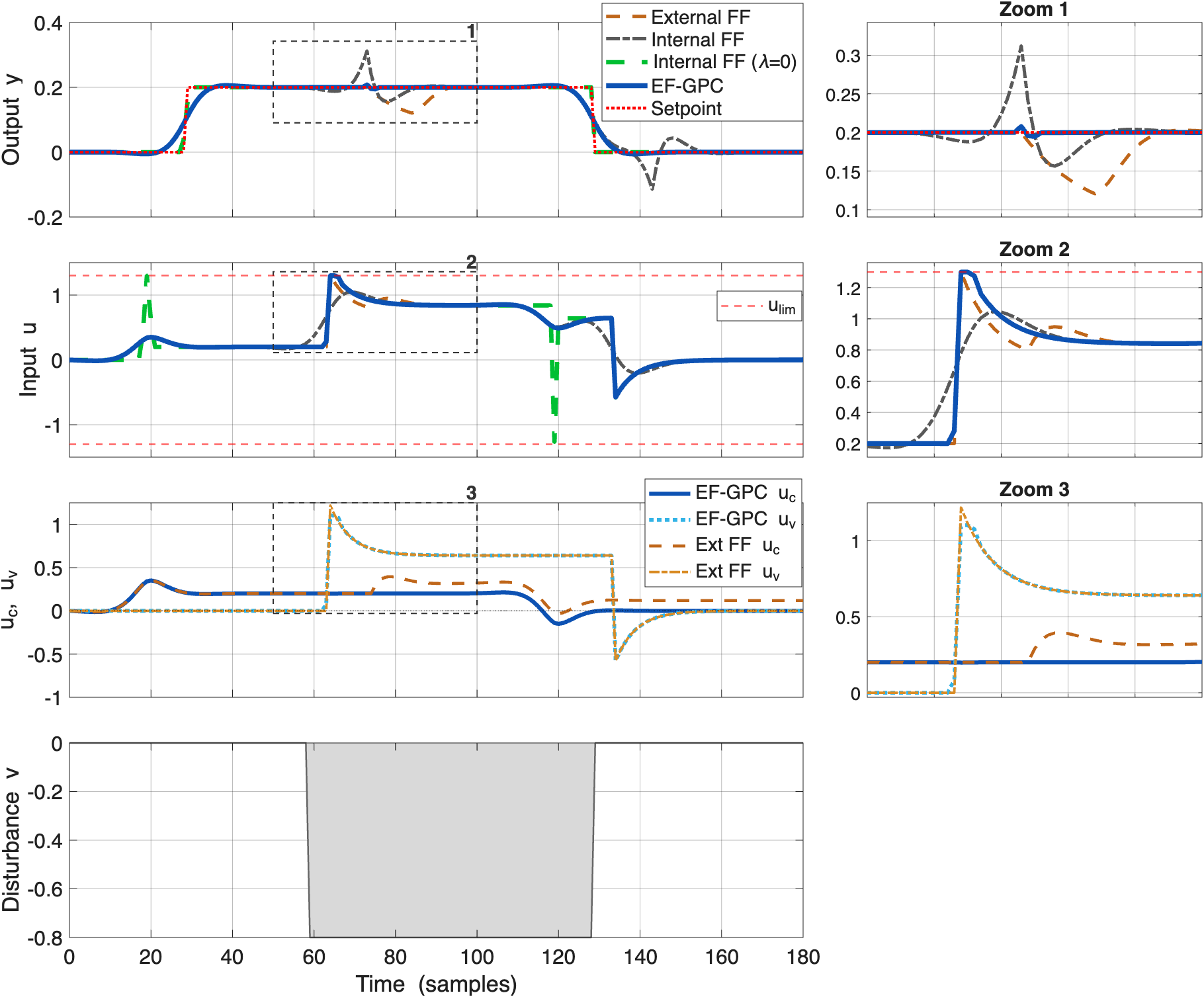}
    \caption{Comparative simulation results for the input-constrained feedforward compensation scenario}
    \label{fig:constraint}
\end{figure}

The results obtained from the corresponding simulation study are presented in Figure~\ref{fig:constraint}. In this second scenario, a larger set-point change was introduced in comparison with the previous case in order to establish an operating point close to input saturation. Despite this modification, the overall closed-loop responses remain qualitatively similar to those obtained in the ideal unconstrained scenario. 

The main differences appear during the measurable disturbance rejection stage, particularly when comparing the proposed EF-GPC formulation with the classical External FF strategy. As can be observed, the proposed EF-GPC approach achieves a fast disturbance rejection (Zoom 1) while properly handling the input saturation constraints (Zoom 2), taking advantage of the inherent capability of predictive controllers to explicitly manage constrained control actions. In contrast, in the External FF strategy, the feedforward compensator drives the control action into saturation, after which the feedback controller must subsequently compensate for the remaining error (Zoom 3). As a consequence, both control contributions operate in a completely decoupled manner, generating conflicting control actions. This behaviour can be clearly appreciated in the figure showing the individual contributions of $u_c$ and $u_v$. The same figure illustrates that, in the proposed EF-GPC formulation, the disturbance rejection task is entirely handled by the feedforward-oriented contribution $u_v$, while the feedback-oriented component preserves the nominal tracking behaviour (see Zoom 3). Finally, during the simultaneous disturbance and set-point change scenario, similar results to those obtained in the previous case were observed.

The quantitative performance indices associated with the constrained scenario are summarized in Table~\ref{tab:ff_comparison_constrained}. The obtained results confirm the conclusions previously observed in Figure~\ref{fig:constraint}. In particular, the proposed EF-GPC formulation achieves a significantly lower disturbance rejection error than the classical External FF strategy, reducing IAE$_{\mathrm{Dis}}$ from 0.264 to 0.020 while preserving an identical tracking performance. Consequently, the overall performance index IAE$_{\mathrm{Tot}}$ is also notably improved. These results clearly illustrate the advantage of integrating the feedforward compensation mechanism directly within the predictive control framework, allowing the controller to explicitly coordinate the disturbance rejection and tracking objectives under constrained operating conditions. 

In contrast, the classical External FF strategy suffers from the decoupled nature of the feedforward and feedback actions, leading to input saturation and a subsequent deterioration of the disturbance rejection performance. Similarly, the conventional Internal FF formulation exhibits the worst overall performance due to the compromise imposed by the use of a single tuning parameter for both objectives. 

\begin{table}[ht]
\centering
\caption{Comparison of performance indices for the input-constrained feedforward compensation scenario.}
\label{tab:ff_comparison_constrained}

\renewcommand{\arraystretch}{1.2}

\begin{tabular}{lcccc}
\toprule
\textbf{Strategy} &
\textbf{IAE$_{\mathrm{Ref}}$} &
\textbf{IAE$_{\mathrm{Dis}}$} &
\textbf{IAE$_{\mathrm{Mix}}$} &
\textbf{IAE$_{\mathrm{Tot}}$} \\
\midrule
EF-GPC                      & 0.206 & 0.020 & 0.206 & 0.431 \\
External FF                 & 0.206 & 0.264 & 0.206 & 0.675 \\
Internal FF                 & 0.213 & 0.572 & 1.635 & 1.635 \\
Internal FF ($\lambda=0$)   & 0.000 & 0.020 & 0.000 & 0.020 \\
\bottomrule
\end{tabular}

\end{table}

These results demonstrate that the proposed EF-GPC formulation is capable of completely decoupling the feedback-oriented and feedforward-oriented control actions, while simultaneously exploiting the inherent advantages of predictive controllers for explicit constraint handling in comparison with classical external feedforward compensation schemes. Moreover, the proposed strategy achieves the same ideal disturbance rejection performance as the classical Internal FF ($\lambda = 0$), while preserving an independent and non-aggressive tuning for the set-point tracking objective. This feature constitutes one of the main advantages of the proposed methodology, since it allows both control objectives to be addressed independently without sacrificing nominal closed-loop behaviour or disturbance rejection capabilities.

\subsubsection{Non-invertible feedforward compensation scenario}

\begin{figure}[h]
    \centering
\includegraphics[width=\linewidth]{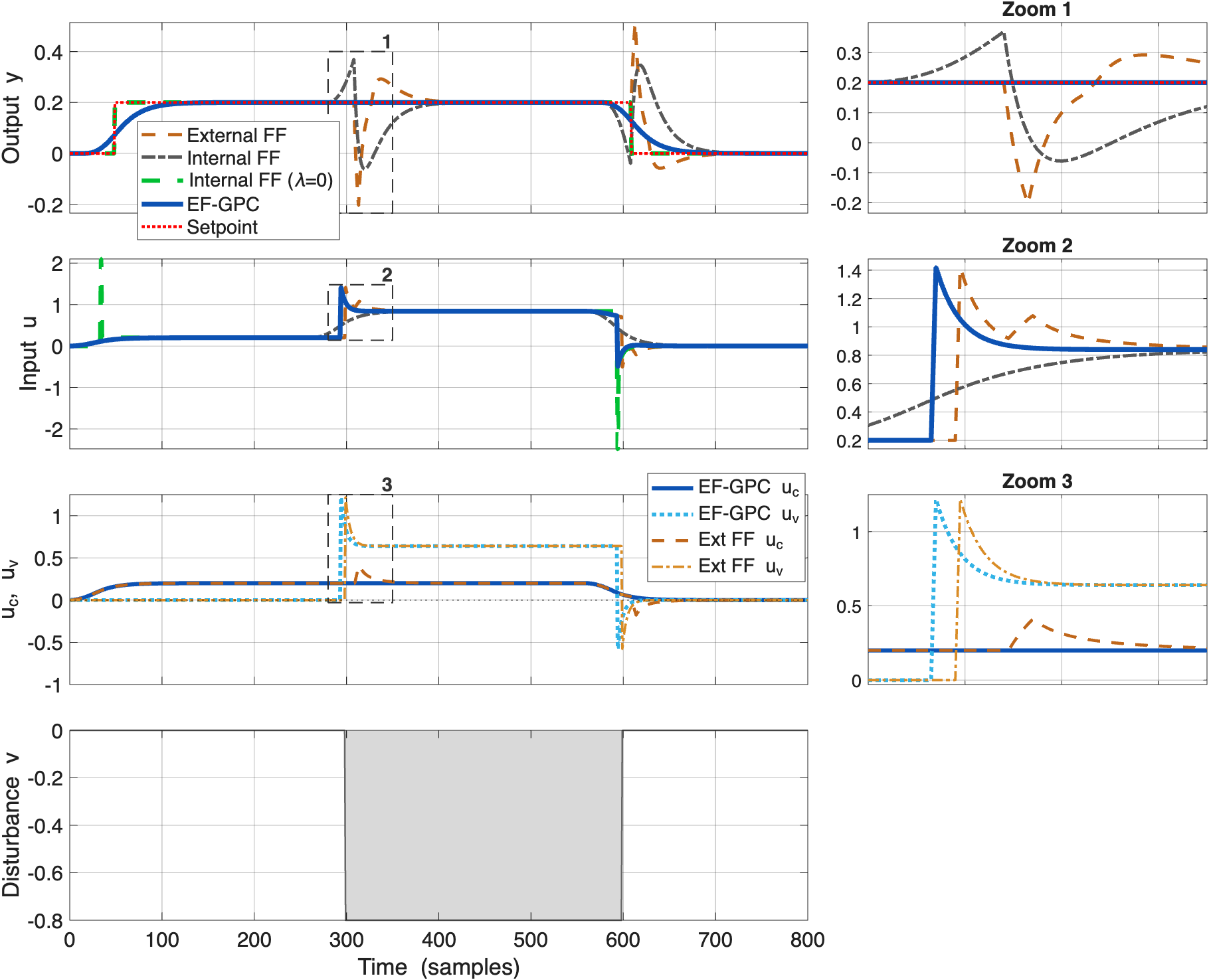}
    \caption{Comparative simulation results for the non-invertible feedforward compensation scenario}
    \label{fig:NInver}
\end{figure}

The third considered case corresponds to a non-invertible feedforward compensation scenario, where perfect disturbance rejection is no longer theoretically achievable. In this case, the process and disturbance dynamics are given by:
\begin{align}
    P_u(s) &= \frac{1}{10s + 1} e^{-15s}, \\
    P_v(s) &= \frac{0.8}{5s + 1} e^{-10s},
\end{align}

\noindent where the manipulated-input transfer function exhibits a larger delay than the disturbance transfer function. Consequently, the disturbance affects the process output before the manipulated variable can compensate for its effect, leading to a non-causal feedforward compensator according to classical feedforward control theory. The results obtained from the corresponding simulation study are presented in Figure~\ref{fig:NInver}, where no constraints were considered in this case.

In this scenario, the set-point tracking responses are very similar to those obtained in the first considered case. The main differences again appear during the disturbance rejection stage. In this case, the classical external feedforward compensator cannot be fully implemented to achieve perfect disturbance cancellation due to the non-causal nature of the required inversion. Consequently, the External FF strategy was implemented neglecting the delay mismatch between the process and disturbance dynamics. As a result, the feedforward action is no longer capable of completely rejecting the measurable disturbance (Zoom 1) and, therefore, the feedback controller must subsequently compensate for the remaining error (Zoom 3). This behaviour causes the feedforward and feedback actions in the External FF strategy to operate independently, as can be clearly observed in the figure showing the individual contributions of $u_c$ and $u_v$ (Zoom 3).

In contrast, in the proposed EF-GPC formulation, the inversion-related limitations are intrinsically handled within the internal predictive model of the controller. As can be observed, the feedforward-oriented contribution $u_v$ is capable of entirely compensating the measurable disturbance (Zoom 3), achieving a disturbance rejection performance identical to the ideal case represented by the  Internal FF ($\lambda=0$) strategy (see Zoom 1), while simultaneously preserving the tracking-oriented component fully decoupled from the disturbance rejection task. Finally, when a simultaneous set-point change and measurable disturbance are introduced, similar limitations can again be observed in the classical External FF strategy, whereas the proposed formulation preserves a significantly better coordinated closed-loop response.

The quantitative performance indices reported in Table~\ref{tab:ff_comparison_noninvertible} further confirm the conclusions extracted from the dynamic responses. In particular, the proposed EF-GPC formulation achieves the same set-point tracking performance as the classical External FF strategy, while substantially improving the disturbance rejection capability under non-invertible conditions. Specifically, the proposed methodology reduces IAE$_{\mathrm{Dis}}$ from 2.749 to 0 and decreases the overall performance index IAE$_{\mathrm{Tot}}$ from 10.69 to 0.412, corresponding to a reduction of approximately 96\%. These results clearly demonstrate the capability of the proposed predictive formulation to properly coordinate the feedforward and feedback actions while intrinsically handling the limitations associated with non-invertible feedforward compensation.

In contrast, both the classical External FF and the conventional Internal FF strategies exhibit a significant deterioration in the disturbance rejection performance due to the delay mismatch between the manipulated-input and disturbance dynamics and the coupled tuning, respectively. Finally, although the Internal FF ($\lambda=0$) strategy theoretically achieves perfect rejection, such behaviour is again obtained at the expense of highly aggressive and unrealistic control actions. 

\begin{table}[h]
\centering
\caption{Comparison of performance indices for the non-invertible feedforward compensation scenario}
\label{tab:ff_comparison_noninvertible}

\renewcommand{\arraystretch}{1.2}

\begin{tabular}{lcccc}
\toprule
\textbf{Strategy} &
\textbf{IAE$_{\mathrm{Ref}}$} &
\textbf{IAE$_{\mathrm{Dis}}$} &
\textbf{IAE$_{\mathrm{Mix}}$} &
\textbf{IAE$_{\mathrm{Tot}}$} \\
\midrule
EF-GPC                    & 0.206 & 0.000 & 0.206 & 0.412 \\
External FF               & 0.206 & 2.749 & 7.734 & 10.690 \\
Internal FF               & 0.256 & 0.597 & 0.777 & 1.629 \\
Internal FF ($\lambda=0$) & 0.000 & 0.000 & 0.000 & 0.000 \\
\bottomrule
\end{tabular}

\end{table}

\subsection{Reverse Osmosis case study}
\begin{figure}[h]
    \centering
\includegraphics[width=\linewidth]{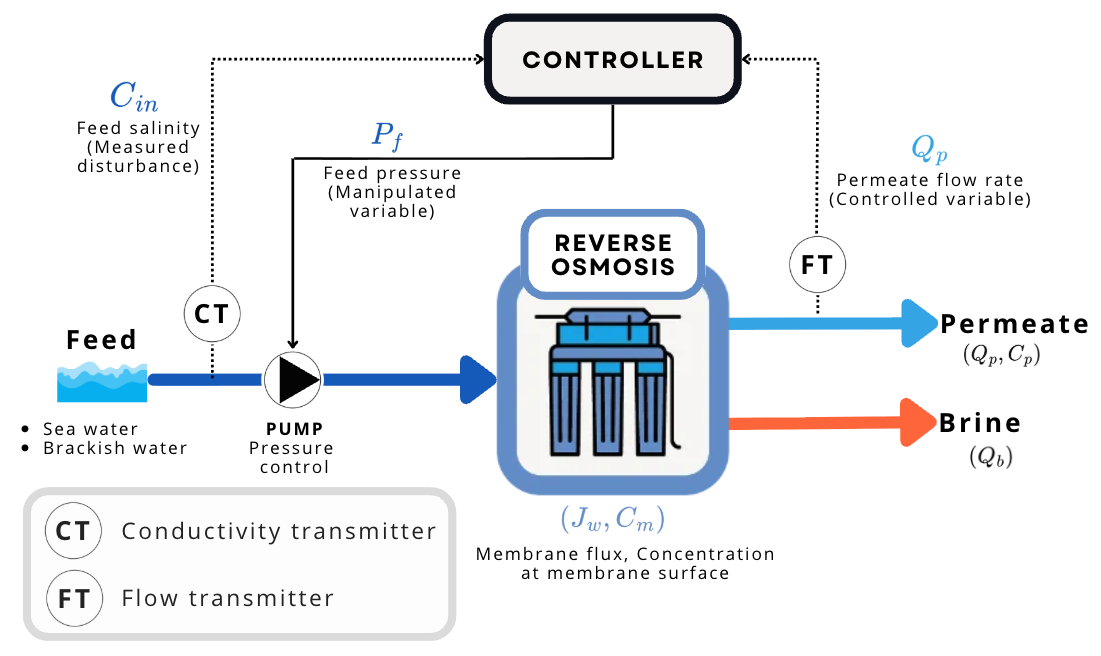}
    \caption{Schematic diagram of the RO system.}
    \label{fig:RO}
\end{figure}

To further assess the applicability of the proposed embedded feedforward formulation beyond the nominal linear examples, the EF-GPC strategy is evaluated in a nonlinear process simulation. The selected process is a reverse osmosis (RO) unit, a membrane-based separation process in which a pressurised saline feed stream is forced through a semi-permeable membrane. Water permeates through the membrane, whereas most of the dissolved salt is retained in the brine stream. The manipulated variable is the feed pressure, $P_f$, and the measured disturbance is the feed salinity, $C_{in}$. The controlled output is the permeate flow rate, $Q_p$, expressed in $\mathrm{m^3/h}$. In this process, variations in feed salinity modify the osmotic pressure and directly affect the net driving force through the membrane, making measurable disturbance compensation especially relevant. A schematic representation of the reverse-osmosis system and the associated control structure is shown in Figure~\ref{fig:RO}.

The nonlinear reverse-osmosis model used to simulate the plant is based on the classical solution–diffusion transport mechanism commonly employed in reverse-osmosis modelling. Similar dynamic formulations can be found in the literature \citep{zhao2011steady}. The resulting low-order model is described by:
\begin{align}
\frac{Q_p(t)}{dt} &= \frac{-Q_p(t) + A_m J_w(t)}{\tau_p}, \\
\frac{C_m(t)}{dt} &= \frac{Q_f C_{in}(t)-Q_b(t) C_m(t)-Q_p(t) C_p(t)}{V},
\label{eq:ro_nonlinear_model}
\end{align}
\noindent in which
\begin{align}
    J_w(t) &= L_p\left(P_f(t)-P_p-k_{\pi}\big(C_m(t)-C_p^\star(t)\big)\right),\\
    C_p(t) &= \frac{B_s}{J_w(t)+B_s}C_m(t), \\
    Q_b(t) &= Q_f-Q_p(t),
\end{align}
where $C_m(t)$ is the average membrane salinity, $C_p(t)$ is the permeate salinity, $J_w(t)$ is the water flux, and $Q_b(t)$ is the brine flow rate. The osmotic-pressure term is computed using $C_p^\star(t)=0.02\,C_m(t)$. The description and values of the model parameters employed in the simulations are summarized in Table~\ref{tab:ro_parameters}. 

\begin{table}[ht]
\centering
\caption{Parameters of the nonlinear reverse-osmosis model used in simulation.}
\label{tab:ro_parameters}

\renewcommand{\arraystretch}{1.15}

\begin{tabular}{ccc}
\toprule
\textbf{Parameter} & \textbf{Value} & \textbf{Description} \\
\midrule
$V$         & $0.02~\mathrm{m^3}$                    & Tank volume \\
$A_m$       & $8~\mathrm{m^2}$                       & Membrane area \\
$L_p$       & $4.0\times10^{-7}~\mathrm{m/(s\,bar)}$ & Water permeability coefficient \\
$B_s$       & $2.0\times10^{-6}~\mathrm{m/s}$        & Salt permeability coefficient \\
$k_{\pi}$   & $0.06~\mathrm{bar/(g/L)}$              & Osmotic-pressure coefficient \\
$P_p$       & $1~\mathrm{bar}$                       & Permeate pressure \\
$Q_f$       & $1.2\times10^{-4}~\mathrm{m^3/s}$      & Feed flow rate \\
$\tau_p$    & $400~\mathrm{s}$                       & Permeate flow time constant \\
\bottomrule
\end{tabular}

\end{table}

The GPC controllers are formulated from the linear model obtained by linearising the nonlinear plant around the nominal operating point given by Table~\ref{tab:ro_parameters}. The transfer function from the manipulated variable to the controlled output is
\begin{equation}
P_u(s)= \frac{Q_p(s)}{P_f(s)} = 
\frac{ 10.1 \times 10^{-3} (784.74s+1)}
{2.7617\times 10^{5}s^2+1.0478\times 10^{3}s+1}.
\label{eq:ro_gu}
\end{equation}
Similarly, the transfer function from the measured disturbance to the controlled output is
\begin{equation}
P_v(s)= \frac{Q_p(s)}{C_{in}(s)} = 
\frac{2.8 \times 10^{-3}}
{2.7617\times 10^{5}s^2+1.0478\times 10^{3}s+1}.
\label{eq:ro_gd}
\end{equation}

Simulation considered a sample time of $T_s = 60$ s, and the horizons $N = 30$, and  $N_u = 12$, based on the equivalent system time-constant $\tau_{eq} = 527$ s. Similar to Section \ref{sec:simulation}, the goal is to compare four predictive strategies: a standard GPC without feedforward action, the Internal FF GPC formulations with different values of $\lambda$ ($\lambda = 0$ and $\lambda=1$), and the proposed EF-GPC with $\lambda_c = 1$ and $\lambda_v = 0$. 

\begin{figure}[!h]
    \centering
    \includegraphics[
        width=\linewidth,
        trim={0 0 0 0.7cm},
        clip
    ]{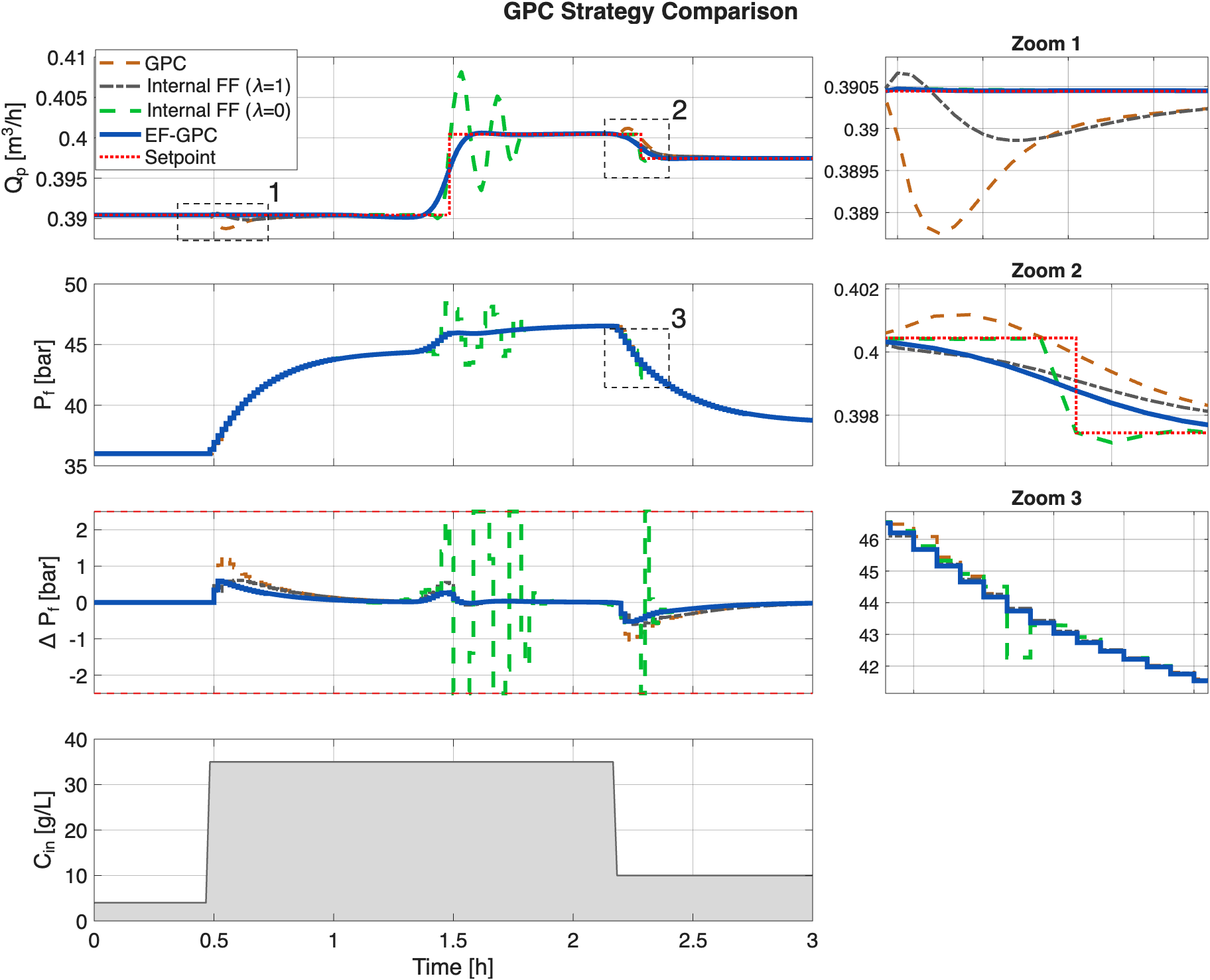}
    \caption{Comparison of the RO nonlinear case study under disturbance and reference changes.}
    \label{fig:ro_case_study}
\end{figure}

The results of the simulation are shown in Figure~\ref{fig:ro_case_study}.  In the first interval (Zoom 1), a large salinity disturbance (see grey signal at the bottom plot of Figure~\ref{fig:ro_case_study}) is applied at the linearisation operating point. In this case, both EF-GPC and the internal feedforward strategy with $\lambda=0$ almost completely reject the disturbance, as expected from their more aggressive disturbance-compensation behaviour. However, the main drawback of the $\lambda=0$ strategy becomes evident when a reference change is introduced (Zoom 2). Since the input component is no longer attenuated, the resulting control action becomes excessively aggressive, surpassing the imposed $\Delta u$ limit and producing a highly oscillatory output response. In contrast, EF-GPC and the other regularised strategies provide smoother closed-loop behaviour during the reference-tracking interval.

The last intervention considers a disturbance and a reference change occurring simultaneously (Zoom 3). Under this more demanding condition, EF-GPC preserves a response very similar to that obtained during the pure reference-change test, indicating that the measurable disturbance is effectively rejected while the tracking task is maintained. The Internal FF with $\lambda=0$ strategy is also almost perfect in terms of disturbance rejection in this interval, but its previous oscillatory behaviour highlights its lack of robustness as a general tuning choice. As expected, the standard GPC and the Internal FF GPC with $\lambda=1$ are slower in this interval because they do not fully compensate for the measurable disturbance.

Table~\ref{tab:controller_comparison} summarises the IAE results obtained for the evaluated strategies. EF-GPC achieved the best overall performance, reducing the total IAE by 43.6\% compared to conventional GPC and by 33.6\% relative to the Internal FF strategy with $\lambda=1$. For the disturbance-rejection interval, EF-GPC and the Internal FF with $\lambda=0$ obtained almost complete rejection (0.0002). However, during the reference-tracking interval, the Internal FF with $\lambda=0$ strategy produced the worst performance (IAE = 0.0709) due to the oscillatory behaviour caused by input increment limits. On the other hand, the EF-GPC preserved the same tracking performance as conventional GPC and the Internal FF with $\lambda=1$ formulation (IAE $\approx$ 0.0232). 

In the combined disturbance and reference-change interval, EF-GPC reduced the IAE by 48.9\% and 42.9\% relative to conventional GPC and the $\lambda=1$ strategy, respectively. These results show that EF-GPC preserves the strong disturbance-rejection capability of the aggressive $\lambda=0$ feedforward formulation while maintaining the smooth reference-tracking behaviour of the regularised predictive controllers.

\begin{table}[!h]
\centering
\caption{Comparison of performance indices for different controllers under disturbance (Dist) and set-point tracking (Track) scenarios.}
\label{tab:controller_comparison}

\renewcommand{\arraystretch}{1.2}

\begin{tabular}{lcccc}
\toprule
\textbf{Strategy} &
\textbf{IAE$_{\mathrm{Ref}}$} &
\textbf{IAE$_{\mathrm{Dis}}$} &
\textbf{IAE$_{\mathrm{Mix}}$} &
\textbf{IAE$_{\mathrm{Tot}}$} \\
\midrule
GPC (no FF)                        & 0.0169 & 0.0232 & 0.0141 & 0.0621 \\
Internal FF ($\lambda=1$)   & 0.0087 & 0.0232 & 0.0126& 0.0527  \\
Internal FF ($\lambda=0$)   & 0.0002 & 0.0709 & 0.0007& 0.0724  \\
EF-GPC ($\lambda_c = 1$, $\lambda_v = 0$)                    & 0.0002 & 0.0232 & 0.0072& 0.0350  \\
\bottomrule
\end{tabular}
\end{table}

\section{Conclusions}       
\label{sec:conclusions}

This work has presented a novel methodology for systematically embedding feedforward capabilities into MPC algorithms. Although measurable disturbances can be explicitly incorporated into standard MPC prediction models, complete disturbance rejection cannot generally be achieved because the control effort penalty inherently limits the aggressive control actions required for perfect compensation. As a consequence, conventional MPC formulations are unable to fully reproduce the behaviour of classical feedforward compensators.

The proposed EF-GPC decompose the control action into two complementary components. The first component preserves the conventional MPC objectives associated with set-point tracking, robustness, and constraint handling, while the second component is specifically devoted to measurable disturbance rejection and is formulated without penalising control effort. This structure enables the recovery of the anticipative behaviour of classical feedforward compensation while retaining the optimisation and constraint-handling capabilities of MPC.

An important advantage of the proposed methodology is that feedforward and feedback objectives can be designed independently and subsequently coordinated through a single optimisation framework. Unlike classical feedforward schemes, actuator saturation, rate constraints, and other standard MPC constraints can be systematically imposed on the combined control action, allowing the interaction between the two components to be naturally managed by the optimiser.

The simulation results showed that the feedforward-oriented predictive action can recover the disturbance-compensation performance of classical feedforward controllers while preserving the desirable closed-loop properties of MPC. Furthermore, the reverse osmosis case study demonstrated the methodology's practical applicability in a realistic process control scenario, highlighting its ability to enhance disturbance rejection without compromising tracking performance or constraint satisfaction.

The developments presented in this work have intentionally focused on the nominal case to clearly establish the proposed methodology and to analyse its feedforward properties. Future research will address robustness and stability aspects in the presence of model uncertainty, disturbance prediction errors, and plant-model mismatch. 

\section*{Acknowledgments}
This work has been financed by the following projects: PID2023-150739OB-I00 and PDC2025-165379-I00 financed by the Spanish Ministry of Science.        

\bibliographystyle{elsarticle-num-names} 
\bibliography{Biblio}

\end{document}